\documentclass[12pt,noshowpacs,amsmath,amssymb,pre,endfloats]{revtex4}
\usepackage{epsfig} 
\usepackage{amssymb} 

\newcommand{\fig}[1]{Figure~\ref{#1}}

\usepackage{graphicx,amsmath,epsfig}
\usepackage{amsfonts}
\usepackage{amssymb}
\usepackage{amsthm}

\begin{document}
\preprint{}
\title{The nonequilibrium Ehrenfest gas: a chaotic model with flat obstacles?}
\author{Carlo Bianca}
\email{carlo.bianca@polito.it}
\affiliation{Dipartimento di Matematica ed Informatica, Universit\'a di Catania, Viale Andrea Doria 6, 95125 Catania, Italy}
\author{Lamberto Rondoni}
\email{lamberto.rondoni@polito.it}
\affiliation{Dipartimento di Matematica and CNISM, Politecnico di
  Torino, Corso Duca degli Abruzzi 24, 10129 Torino, Italy}
\date{\today}

\begin{abstract}

It is known that the non-equilibrium version of the Lorentz gas
(a billiard with dispersing obstacles  \textbf{\cite{Sin70}}, electric field and Gaussian
thermostat) is hyperbolic if the field is small  \textbf{\cite{CHE}}. Differently the
hyperbolicity of the non-equilibrium Ehrenfest gas constitutes an open problem,
since its obstacles are rhombi and the techniques so far developed rely on the
dispersing nature of the obstacles \textbf{\cite{CHE,Wo1}}. We have developed
analytical and numerical investigations which support the idea that this model
of transport of matter has both chaotic (positive Lyapunov exponent)
and non-chaotic steady states with a quite peculiar sensitive dependence
on the field and on the geometry, not observed before. The associated
transport behaviour is correspondingly highly irregular, with features whose
understanding is of both theoretical and technological interest.

\begin{sloppypar}
{\it AMS}:  82C05, 70Fxx, 82C70
\end{sloppypar}
\end{abstract}

\maketitle


\section{Introduction}
The Ehrenfest model of diffusion (named after the Austrian Dutch
physicists Paul and Tatiana Ehrenfest) was proposed in the early
1900s in order to illuminate the statistical interpretation of the
second law of thermodynamics and to study the applicability of the Boltzmann
equation. In the Ehrenfest wind-tree model \textbf{\cite{EHE}},
the point-like (''wind'') particles move on a plane and collide
with randomly placed fixed square scatterers (''tree'').\\
This model has been recently reconsidered in \textbf{\cite{DETT}}
to prove that microscopic chaos is not necessary for Brownian
motion. A one-dimensional version of this model has been considered in \textbf{\cite{CECC}} to investigate the origin of diffusion in non chaotic systems. In   \textbf{\cite{CECC}} the authors identify two sufficient ingredients for diffusive behavior in one-dimensional,
non-chaotic systems: a) a finite-size, algebraic instability mechanism, and b) a mechanism that
suppresses periodic orbits. A  nonequilibrium modification of the model, with regularly
placed scatterers, has been proposed in
\textbf{\cite{LRB}} to test the applicability of the so called
fluctuation relation (\textbf{\cite{ECM2,ES,GC1,GC2}}) to non chaotic systems.
This modified model was chosen under the assumption that, at small and vanishing
fields at least, it must be non-chaotic, since
collisions with flat boundaries do not lead to exponential
separation of nearby phase space trajectories.\\
However the question of whether such a model can have positive
Lyapunov exponents, as functions of the field, is open. Indeed, the
techniques so far developed, e.g.\ by Chernov
and Wojtkowski \textbf{\cite{CHE,Wo1}},  rely on the
dispersing nature of the billiard obstacles. \\
In this paper, the
dynamical properties of the nonequilibrium version of the
Ehrenfest gas are considered as functions of the field and of the parameters which
determine the billiard table. Numerical tests are performed to
find chaotic attractors and to compute the Lyapunov exponents. The construction of a sort of
bifurcation diagram of the attractor as a function of the electric
field and of the geometry is attempted. The result turns out to be
quite peculiar: chaotic regimes with an extremely sensitive dependence on the
parameters appear possible, although not easy to establish rigorously.
If this model can be taken to approximate the transport of matter in microporous
membranes, our results confirm the sensitive dependence of microporous transport
on all relevant parameters observed e.g.\ in \textbf{\cite{Bia,Je-Ro}}.
Indeed, the current of the nonequilibrium Eherenfest gas is proportional
to the sum of the Lyapunov exponents (cf. Eq.\eqref{formula} below), which varies with
the parameters as irregularly as the attractors do.


\section{The model}

\begin{center}
\begin{figure}
\epsfig{file=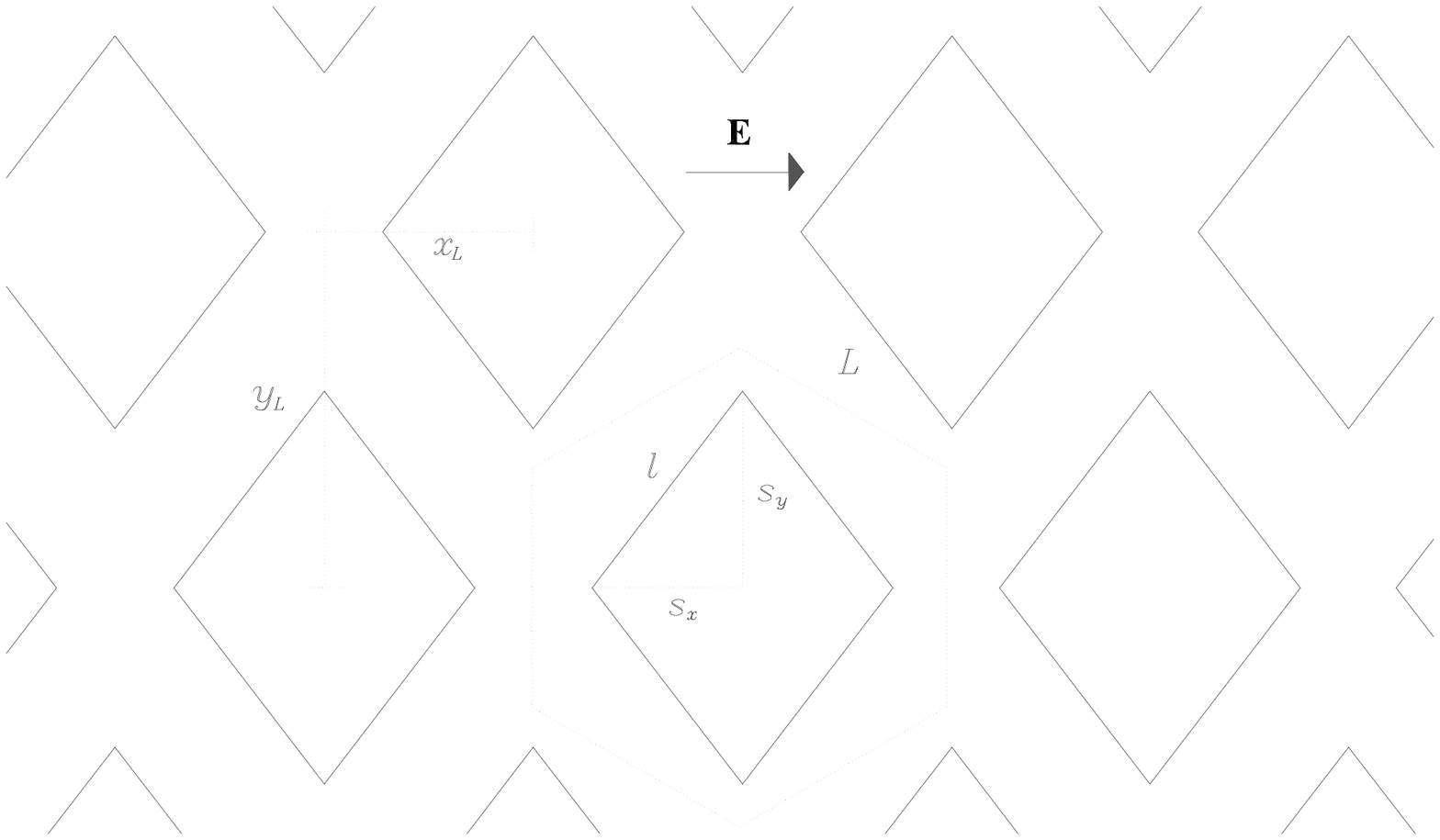,angle=0,width=8cm} \caption{The
modified Ehrenfest gas. In the present paper, the side of the
elementary cell is set to 1.291, while the semiaxis of the rhombus
are chosen to be 1.1 and 0.7573 respectively.} \label{elec}
\end{figure}
\end{center}

The billiard table consists of rhombi of side length $l$ with
distances along the $x$ and $y$ directions between the centers of
two nearest neighbouring rhombi given by $x_{L}$ and $y_{L}$,
respectively. The centers of the rhombi are fixed on a triangular
lattice in a plane and have coordinates

$$\left(%
\begin{array}{c}
  x_{c} \\
  y_{c} \\
\end{array}%
\right)=m_{c}\textbf{l}_{1}+n_{c}\textbf{l}_{2}, \qquad m_{c},
n_{c} \in \mathbb{Z}$$ where $\textbf{l}_{1}=(x_{L},0)$ and
$\textbf{l}_{2}=(0, y_{L})$ are the lattice vectors. If all the
pairs $(m_{c},n_{c})$ are selected, the billiard is invariant
under the group of spatial translations generated by
$\textbf{l}_{1}$ and $\textbf{l}_{2}$. Accordingly, the whole
lattice can be mapped onto a so-called Wigner-Seitz cell, with
periodic boundary conditions (\fig{elec}). The elementary
Wigner-Seitz cell of the triangular lattice is a hexagon of length
side $L$ and area
$$A_{WS}=|\textbf{l}_{1}\times \textbf{l}_{2}|.$$
The centers of all other cells are identified by the pairs
$(m_{c},n_{c})\in \mathbb{Z}^{E} \times \mathbb{Z}^{E}$ or
$(m_{c},n_{c})\in \mathbb{Z}^{O} \times \mathbb{Z}^{O}$ where
$\mathbb{Z}^{E}=\{n \in \mathbb{Z}: |n| \, \textrm{is even}\},
\quad \mathbb{Z}^{O}=\{n \in \mathbb{Z}: |n| \, \textrm{is
odd}\}.$
Because of the bi--jective correspondence between rhombi and pairs
$(m_{c},n_{c})$, one may label a generic rhombus by
$\mathcal{R}_{m_{c},n_{c}}$ and the corresponding hexagon by
$\mathcal{H}_{m_{c},n_{c}}$. Further, a label can be put on the
sides of the rhombi and of the hexagons, introducing an alphabet
$\mathcal{A}=\{r_{1},r_{2},r_{3},r_{4}\}$, starting from the right
vertical side and oriented clockwise, for the sides of the rhombi
and an alphabet $\mathcal{B}=\{h_{1}, h_{2},
h_{3},h_{4},h_{5},h_{6}\}$ for the hexagon sides, starting from
the right vertical side and oriented clockwise. In this alphabet,
the sides of a generic rhombus of the lattice can be labelled by a
triple $(\mathcal{R}_{m_{c},n_{c}}, s)$ with $s\in \mathcal{A}$,
while the hexagon side with a triple $(\mathcal{H}_{m_{c},n_{c}},
s)=(m_{c},n_{c},s)$ with $s\in \mathcal{B}$. The rhombi lying in
the $y$ axes have $\mathcal{R}_{0,i}=(0,i)$ with $i \in
\mathbb{Z}^{O}$ and the ones lying in the $x$ axes have
$\mathcal{R}_{j,0}=(j,0)$ with $j \in \mathbb{Z}^{E}$.\\
The geometry of the model is determined by the side $L$ of the
hexagonal Wigner-Seitz cell, so that $x_{L}=\sqrt{3}/{2}$
and $y_{L}=3L/2$. Let $\mathcal{R}_{0}=(0,0)$ be the
rhombus with the center in the origin of the Cartesian
coordinates, $l$ its sides length, $s_{x}$ and $s_{y}$ the half
length of the major and minor diagonals respectively, so that
$l=\sqrt{s_{x}^{2}+s_{y}^{2}}$. To prevent the overlap of rhombi,
the side length of the rhombus inside one hexagon has to verify

$$0 \le l \le \dfrac{\sqrt{7}}{2} \,L$$
which implies $0 \le s_{x} \le x_{L}$.
The case with $l=\dfrac{\sqrt{7}}{2} \,L$ corresponds to a billiard table
which was recently considered in \textbf{\cite{Bia, Je-Ro}}.\\
%
Take $l \in \Big(0, \dfrac{\sqrt{7}}{2} \,L \Big )$, hence
$s_{x} < x_{l}$. The horizon of the billiard depends on the
quantity $s_{y}$ and in particular on the difference
$y_{L}-2s_{y}$. If $s_{y} \ge {y_{L}}/{2}$, the horizon is
finite; if $s_{y} < {y_{L}}/{2}$, it is infinite.
The infinite horizon case allows collision-free trajectories,
parallel to the $x-$axis.\\
When the dynamics is followed within to the Wigner-Seitz cell, the
position of the point particle of mass $M$ must be supplemented by
the couple $(m_{c},n_{c})$, in order to determine its actual
position in the infinite plane. The space between the rhombi forms
the two-dimensional domain, in which the particle moves with
velocity $v$ during the free flights, while collisions with the
sides of the rhombi obey the law of elastic
reflection. \\
In order to drive the model out of equilibrium, an external
electric field parallel to the x-axis, $\textbf{E}= \epsilon
\mathbf{\widehat{x}}$ is applied. If there were no interaction
with a thermal reservoir, any moving particle would be
accelerated by the external field, on average, leading to an
indefinite increase of energy in the system, and there would be no
stationary state. Therefore, in \textbf{\cite{LRB}} the particle
has been coupled to a Gaussian thermostat. The resulting model,
with periodically distributed scatterers, has been called the
\textit{non-equilibrium Ehrenfest gas}. Its phase space has
four coordinates $(x, y, p_{x}, p_{y})$ and its equations of motion
are given by
\begin{equation}
\left\{%
\begin{array}{ll}
   \dot{x}={p_{x}}/{M} ~, \qquad \dot{p_{x}}=\epsilon- \alpha(\textbf{p}) p_{x} \\
   \dot{y}={p_{y}}/{M} ~, \qquad \dot{p_{y}}=- \alpha(\textbf{p}) p_{y}
\end{array} \qquad \mbox{with } ~~~~ \alpha(\textbf{p})=- \epsilon p_{x}
\right.
\end{equation}
where $\epsilon$ is the electric field and $\alpha$ the Gaussian thermostat.
The quantity
$-\textrm{div}(\dot{\textbf{p}}, \dot{\textbf{q}})=\alpha=\epsilon p_{x}$
is known as the phase space contraction rate.
Because of the Gaussian thermostat, $p_{x}^{2}+p_{y}^{2}$ is a
constant of motion, hence there are only three independent
variables, and one may replace $p_{x}$ and $p_{y}$ by the  angle
$\theta \in (-\pi, \pi]$ that $\textbf{p}=p(\cos \theta, \sin
\theta)$ forms with the $x$-axis.
For sake of simplicity, we set $M = 1$ and $p=1$.\\
Then, if $(x_{t}, y_{t})$ denotes the position at time $t$ and
$\theta_{t}$ the velocity angle, measured with respect to the
$x$-axis, the trajectory between two collisions reads \textbf{\cite{Ro3}}
\begin{equation}\label{equtraj}
\left\{%
\begin{array}{ll}
    x_{t}=x_{0}-\frac{1}{\epsilon} \ln \frac{\sin \theta_{t}}{\sin \theta_{t_{0}}} \\
    y_{t}=y_{0}- \frac{1}{\epsilon}(\theta_{t}-\theta_{t_{0}}) \\
    \tan \frac{\theta_{t}}{2}= \exp(- \epsilon(t -t_{0})) \tan \frac{\theta_{t_{0}}}{2}
\end{array}%
\right.
\end{equation}
where $t_{0}$ is the time of the previous collision, while the collision
map $C$ is given by
\begin{equation}
\left\{%
\begin{array}{ll}
    x'_{t}=x_{t} \\
    y'_{t}=y_{t} \\
    \theta'_{t}=-\theta_{t} \pm 2 \theta
\end{array}%
\right.
\end{equation}
where $\theta_{t}$ is the incidence angle, $(x_{t}, y_{t})$ is
the collision point and $\theta$ is the angle that the side of the
rhombus makes with the $x-$axis. The sign $\pm$ depends on the
side on which the bounce
occurs. Hence $C$ is piecewise linear in $\theta_{t}$.\\
Considering the dynamics as a geodesic flow on a Riemann manifold,
the appropriate  metric for this system is \textbf{\cite{Wo-Li,DET,MOR}}

$$ds^{2}=e^{-2 \epsilon x} (dx^{2}+dy^{2})$$
which implies that the quantities $\pi_{y}=e^{-\epsilon x }p_{y}$
and $\phi_{t}=\theta_{t}+\epsilon y_{t}=\theta_{t_{0}}+\epsilon
\theta$ are conserved. Also, the path length $\mathcal{L}(P_{0},
P_{t})$ between $P_{0}=(x_{0}, y_{0}, \theta_{t_{0}})$ and
$P_{t}=(x_{t}, y_{t}, \theta_{t})$ turns out to be

\begin{equation}\label{length}
\mathcal{L}(P_{0}, P_{t})= \frac{1}{\epsilon}e^{-\epsilon x_{0}}
\, \sin \theta_{t_{0}}\,|\cot \theta_{t_{0}}- \cot \theta_{t}|.
\end{equation}

\section{Periodic orbits and stability matrices}\label{periodic}

Using the symbols introduced
above, a trajectory segment $\Omega _{N}$ which consists of $N$
collisions can be labelled by a finite symbolic sequence such as
$(\mathcal{R}_{i_{1},j_{1}},s_{1})
(\mathcal{R}_{i_{2},j_{2}},s_{2})\dots
(\mathcal{R}_{i_{N},j_{N}},s_{N})$,
with $(i_{c},j_{c})\in
\mathbb{Z}^{E} \times \mathbb{Z}^{E}$ or $(i_{c},j_{c})\in
\mathbb{Z}^{O} \times \mathbb{Z}^{O}$ and $s_{c} \in \mathcal{A}$
for $c=1,\dots,N$.
Periodic trajectories will be labelled by sequences which are infinitely
many copies of a fundamental finite sequence.\\
There are two types of periodic orbits; those that are periodic in
the plane, i.e. that return to the initial point in the plane
(they are \emph{closed}: $\Delta x_{i}=0$, $\Delta y_{i}=0$) and those
whose periodicity relies upon the periodicity of the triangular
lattice, and return to the same relative position in a different
cell (they are \emph{open}: $\Delta x_{i} \ne 0$ or $\Delta y_{i} \ne 0$).

\begin{figure}
\begin{center}
\epsfig{file=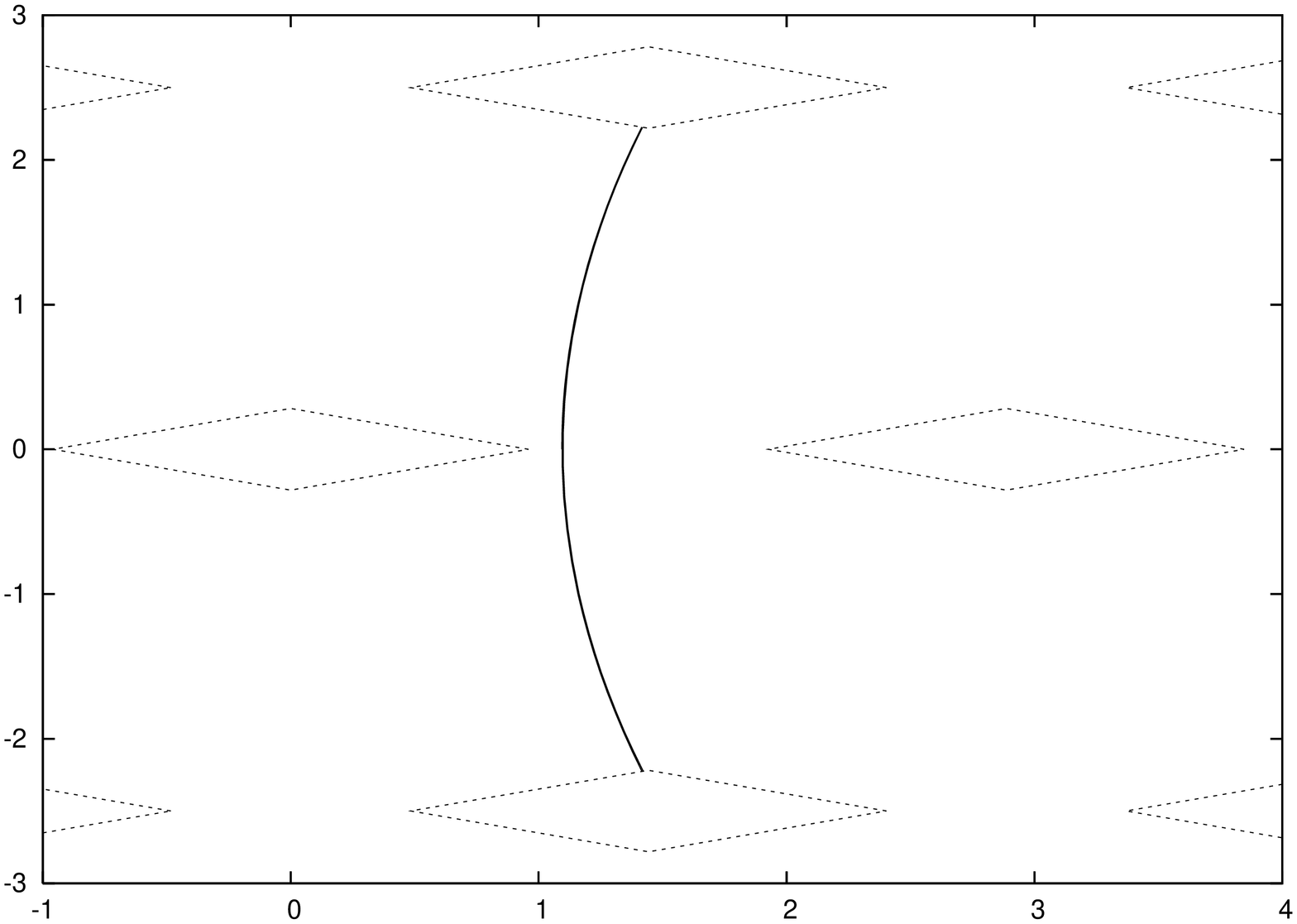,angle=-90,width=8cm}
\epsfig{file=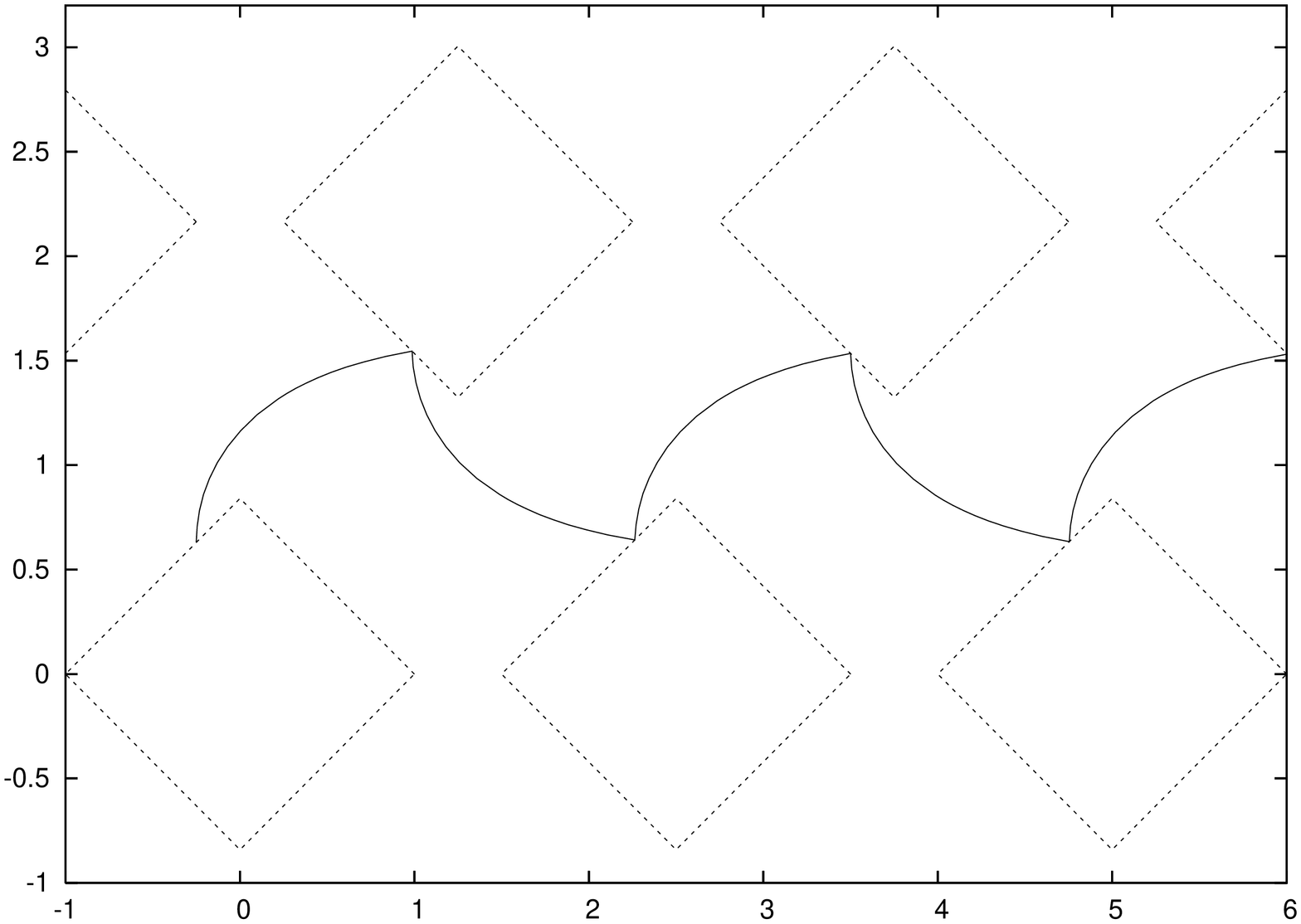,angle=-90,width=8cm}
\caption{The closed
periodic-two orbit $ \Omega^{o}_{2}=(\mathcal{R}_{-1,-1},
r_{4})(\mathcal{R}_{1,1}, r_{3})$
for $\theta=\frac{\pi}{11}$,
$x_{l}=\frac{5\sqrt{3}}{3}$ (left panel). The open
periodic orbit $\Omega^{o}_{2}=(\mathcal{R}_{0,0},
r_{4})(\mathcal{R}_{1,1}, r_{3})(\mathcal{R}_{2,0}, r_{4})$
for $x_{l}=2$, $s_{x}=0.8$, $s_{y}=0.5$ (right panel).} \label{open}
\end{center}
\end{figure}

The velocity vectors of the closed orbits with two collisions
have to be orthogonal to both sides of the rhombi
where collisions occur. This implies $\theta_{0}=\frac{\pi}{2} + \theta$
and $\theta'_{0}=\frac{\pi}{2} - \theta$, where $\theta_{0}$ is the out-going
velocity angle and $\theta'_{0}$ the in-coming angle, because the absolute
value of the velocity angle decreases and preserves the sign during the
free flight \textbf{\cite{Ro3}}.
Then, the closed period-two orbits fly between rhombuses in
the same line parallel to the $y$-axes, and have period $\tau$ and length
$\mathcal{L}$ given by \textbf{\cite{CBthesis}}
$$
\tau=\frac{2}{\epsilon} \ln \frac{\tan \Big( \frac{\pi}{4} + \frac{\theta}{2}\Big)}{\tan \Big( \frac{\pi}{4} - \frac{\theta}{2}\Big)} ~, \qquad
\mathcal{L}=\frac{2}{\epsilon} e^{- \epsilon x_{0}}\sin \theta ~.
$$
Now, fix the geometry of the model through the
parameters $L>0$, $s_{x}>0$, $s_{y}>0$ and take the initial conditions
$$\Big\{x_{0}, \, y_{0}=\pm y_{l}, \,  \theta_{0}=\pm
\dfrac{\pi}{2} \Big\}.$$
It is easy to show that the periodic orbits  $(\mathcal{R}_{0,0},
r_{4})(\mathcal{R}_{0,2}, r_{3})$ (\fig{open}, left panel) and
$(\mathcal{R}_{0,0}, r_{3})(\mathcal{R}_{0,-2}, r_{4})$, with
electric field
\begin{equation}\label{ef1}
 \epsilon= \dfrac{- \theta + \tan \theta \, \ln \cos
\theta}{\mp y_{0}+ m \, x_{0}+ s_{y}} ~,
\end{equation}
exist if and only if \textbf{\cite{CBthesis}}
\begin{equation}\label{cef1}
\dfrac{2 \theta }{3L} <  \epsilon  < \dfrac{2 \theta}{3L- 2s_{y}}.
\end{equation}
%
%
For open orbits with two collisions in the finite horizon case,
simple algebra shows that the following symbolic representations
$(\mathcal{R}_{i,j}, r_{4})(\mathcal{R}_{i-1,j+1},r_{2})(\mathcal{R}_{i+2,j}, r_{4})$
and
$(\mathcal{R}_{i,j}, r_{1})(\mathcal{R}_{i+3,j+1},r_{3})(\mathcal{R}_{i+2,j}, r_{1})$
cannot be realized.
However, orbits with symbolic representation
$(\mathcal{R}_{i,j}, r_{4})(\mathcal{R}_{i+1,j+1},
r_{3})(\mathcal{R}_{i+2,j}, r_{4})$, and its symmetric counterpart
$(\mathcal{R}_{i,j}, r_{3})(\mathcal{R}_{i+1,j-1},
r_{4})(\mathcal{R}_{i+2,j}, r_{3})$ (\fig{open}, right panel) do exist.
Their period is given by
$$\tau=\frac{2}{\epsilon}\,  \ln \frac{1}{\tan (\frac{\pi}{4}-
\theta)}.$$
The open periodic orbits with four collisions (\fig{fig7}, left panel)
and symbolic sequence
$(\mathcal{R}_{0,0},
r_{4})(\mathcal{R}_{-1,1}, r_{2})(\mathcal{R}_{1,1}, r_{3})
(\mathcal{R}_{0,0}, r_{1})(\mathcal{R}_{2,0}, r_{4})$, and
its simmetric image
$(\mathcal{R}_{0,0},
r_{3})(\mathcal{R}_{-1,-1}, r_{1})(\mathcal{R}_{1,-1}, r_{4})
(\mathcal{R}_{0,0}, r_{2})(\mathcal{R}_{2,0}, r_{3})$
do exist if  \textbf{\cite{CBthesis}}
%
%
%
%
\begin{equation}
\max \Big\{-s_{x},-x_{l}+\dfrac{1}{\epsilon} \ln \dfrac{\sin
\theta'_{0}}{\sin \theta_{0}} \Big\}< x_{0}< \min
\Big\{-x_{l}+s_{x}+\dfrac{1}{\epsilon} \ln \dfrac{\sin
\theta'_{0}}{\sin \theta_{0}},0 \Big\}
\end{equation}
%


To compute the Lyapunov exponents for these orbits and any other
trajectory, consider the stability matrix
$J_{S}$ for a trajectory as the product of free flight
stability matrices $J_{F}$ and collision stability matrices
$J_{C}$
$$J_{S}=
\prod_{i=1}^{n}J_{C}(i)J_{F}(i)$$
The number of degrees freedom for the billiard map is two and the
variables that we will use are $(x_{0},\theta_{0})$. Thus
$$J_{C}=
\begin{pmatrix}
 \frac{\partial \theta'}{\partial \theta_{0}} &  \frac{\partial \theta'}{\partial x_{0}}\\
 \frac{\partial x'_{0}}{\partial \theta_{0}} &  \frac{\partial x'_{0}}{\partial x_{0}}.
\end{pmatrix} =
\begin{pmatrix}
 -1 &  0\\
  0 &  1
\end{pmatrix}
$$
The free flight matrix $J_{F}$ depends on the side which the
trajectory leaves and the one which it reaches. There are two
different types of side, the ones with positive slope, of equation
$
y=\tan \theta \, x+ c
$
and the ones with negative slope, of equation
$
y=-\tan \theta \, x+d
$,
where $c$ and $d$ are real numbers.\\
Let us compute the free flight matrix $J_{+,-}$ of a trajectory
which goes from a side with positive slope to a side with negative
slope. Let $(x_{0},y_{0},\theta_{0})$ and
$(x'_{0},y'_{0},\theta'_{0})$ be the initial condition on a side
with positive slope
and the final condition on a side with negative slope respectively.
By using the equations of the trajectory and by the implicit
function theorem \textbf{\cite{CBthesis,IFT}} we obtain
the Jacobian matrix of the free flight:
\begin{equation}\label{jm12}
J_{+,-}=
\begin{pmatrix}
 \dfrac{(\tan \theta_{0}+\tan \theta)\,\tan \theta'_{0}}
{(\tan \theta'_{0}+\tan \theta)\,\tan \theta_{0}} & \dfrac{2 \epsilon\, \tan \theta \, \tan \theta'_{0}}{\tan \theta'_{0}+\tan \theta} \\
& \\
 - \dfrac{1}{\epsilon
\, \tan \theta_{_{0}}}\, \dfrac{ \tan \theta_{0}-\tan
\theta'_{0}}{\tan \theta'_{0}+\tan \theta} & \dfrac{ \tan
\theta'_{0}-\tan \theta}{\tan \theta'_{0}+\tan \theta}
\end{pmatrix}
\end{equation}
Similarly, the flights from a side with negative
slope to a side with positive slope yield
\begin{equation}\label{jm21}
J_{-,+}=
\begin{pmatrix}
 -\dfrac{(\tan \theta_{0}-\tan \theta)\,\tan \theta'_{0}}
{(-\tan \theta'_{0}+\tan \theta)\,\tan \theta_{0}} & \dfrac{2 \epsilon\, \tan \theta \, \tan \theta'_{0}}{-\tan \theta'_{0}+\tan \theta} \\
& \\
- \dfrac{1}{\epsilon \, \tan \theta_{_{0}}}\, \dfrac{ \tan
\theta'_{0}-\tan \theta_{0}}{\tan \theta-\tan \theta'_{0}}  &
\dfrac{ \tan \theta'_{0}+\tan \theta}{\tan \theta'_{0}-\tan
\theta}
\end{pmatrix} ~,
\end{equation}
and those from one side to a parallel one yield
\begin{equation}
J_{\pm , \pm}=
\begin{pmatrix}
  \dfrac{(\tan
\theta_{0}\mp \tan \theta)\,\tan \theta'_{0}} {(\tan
\theta'_{0}\mp \tan \theta)\,\tan \theta_{0}} & 0 \\
&\\
- \dfrac{1}{\epsilon \, \tan \theta_{_{0}}}\, \dfrac{ \tan
\theta_{0} - \tan \theta'_{0}}{\tan \theta'_{0} \mp \tan \theta} & 1
\end{pmatrix} ~.
\end{equation}
If $\mu_{1}$ and $\mu_{2}$ are the eigenvalues
of the stability matrix $J_{S}$ for a periodic orbit of period $\tau$,
the two Lyapunov exponents are $\lambda_{i}=\dfrac{1}{\tau} \log |\mu_{i}|$,
$i=1,2$, and one obtains
\begin{equation}\label{formula}
j = \frac{\Delta x}{\tau}=-\frac{(\lambda_{1}+\lambda_{2})}{\epsilon}
\end{equation}
where $j$ is the current and
$\Delta x$ is the corresponding displacement in the direction of the
field \textbf{\cite{Ro3}}.
Both the Lyapunov exponents of the closed periodic orbits, with period
two, vanish. Indeed, consider that
this periodic orbit has $ \Delta x=0$, hence $\lambda_{1}+\lambda_{2}=0$. Furthermore, the stability matrix of these periodic orbits, which is
$J_{S}=J_{C}J_{-,+}J_{C}J_{+,-}$, is given by
$$J_{S}=
\begin{pmatrix}
 \dfrac{4 \tan^{2} \theta-(1-\tan^{2}\theta)}
{(1+\tan^{2} \theta)^{2}} & \dfrac{4 \epsilon  \tan \theta (\tan^{2}\theta-1)}{(1+\tan^{2} \theta)^{2}} \\
&\\
 \dfrac{4 \epsilon  \tan \theta (\tan^{2}\theta-1)}{(1+\tan^{2} \theta)^{2}}  & -\dfrac{4 \tan^{2} \theta-(1-\tan^{2}\theta)}
{(1+\tan^{2} \theta)^{2}}
\end{pmatrix}
$$
whose determinant is 1, while its trace vanishes. This,
implies that both Lyapunov exponents vanish.
%
%

\section{Numerical estimates of Lyapunov exponents}

In this paper, a system is called chaotic when it has at
least one positive Lyapunov exponent. We note that the boundary of
our system is not defocussing and the external field has a focussing
effect, so the overall dynamics should not be chaotic in
general, although it is not obviuos that this is the case for all
values of the electric field $\epsilon$. In this section we examine the
stationary state and the Lyapunov exponents, obtained by using the algorithm developed by
Benettin, Galgani, Giorgilli and Strelcyn  \textbf{\cite{BGGS}}, for different values
of $\epsilon$, ranging from small to large fields.


\subsection{Chaos for large electric fields}

\begin{figure}
\begin{center}
\epsfig{file=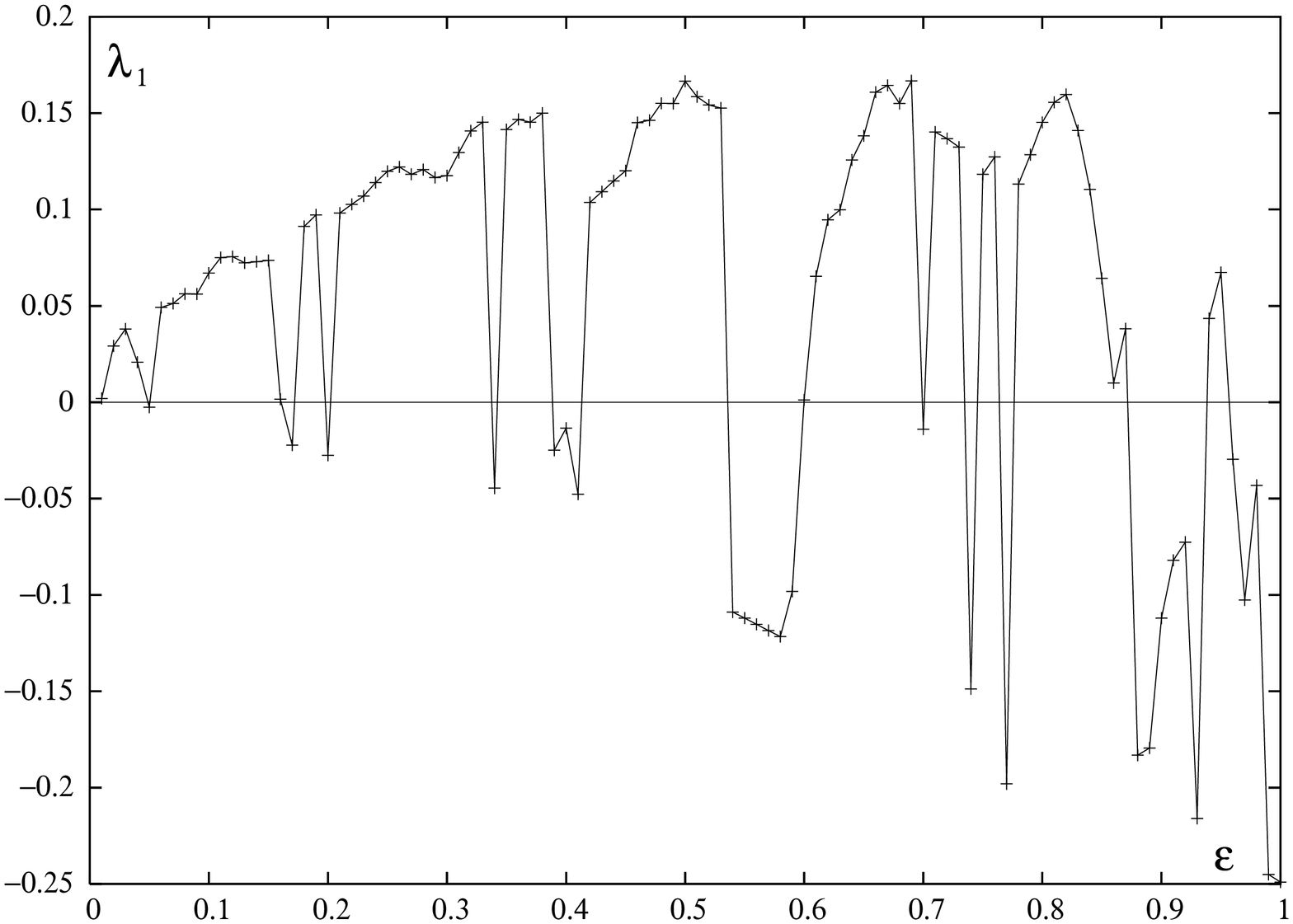,angle=-90,width=10cm} \caption{The largest
Lyapunov exponent $\lambda_1$ for electric fields between $0.02$ and $1$, for
trajectories with $10^{7}$ collisions, and random initial condition.} \label{fig1}
\end{center}
\end{figure}

Numerical simulations of the model starting with random initial
conditions and electric field in the range $[0.02, 1]$ have been
initially performed for a trajectory of length $n=10^{7}$
collisions. The parameters chosen for the geometry are $L=1.291$,
$s_{x}=0.7573$ and $s_{y}=1.1$ which correspond to a case in which
the angles of the rhombi are irrational w.r.t. $\pi$, then,
according to a conjecture by Gutkin \textbf{\cite{GU}}, the equilibrium
version of this model, (i.e. the $\epsilon=0$ case), should be
ergodic. For simulations of $10^{7}$ collisions, \fig{fig1} shows that the fields which
appear to lead to one positive Lyapunov exponent cover a range
larger than that which appears to correspond to two negative
exponents. However, do $10^{7}$ collisions suffice
for a generic trajectory to characterize the stationary state? For
the cases with two negative exponents the answer is affirmative,
since the trajectory is clearly captured by an attracting periodic
orbit. But the cases with one apparently positive exponent are not equally clear. As \textbf{\cite{LRB}} already noted, the doubt is that, starting from
a generic initial condition, convergence to the steady state might be too slow
to be discovered, for reasons which had not been investigated. Indeed, even in cases
in which convergence is observed, the
particle often appears to peregrinate in a sort of
chaotic quasi steady state for very long times, before eventually settling on
a periodic or quasi-periodic steady state. Ref. \textbf{\cite{LRB}}
suggested this might always be the case.

\begin{figure}
\begin{center}
\epsfig{file=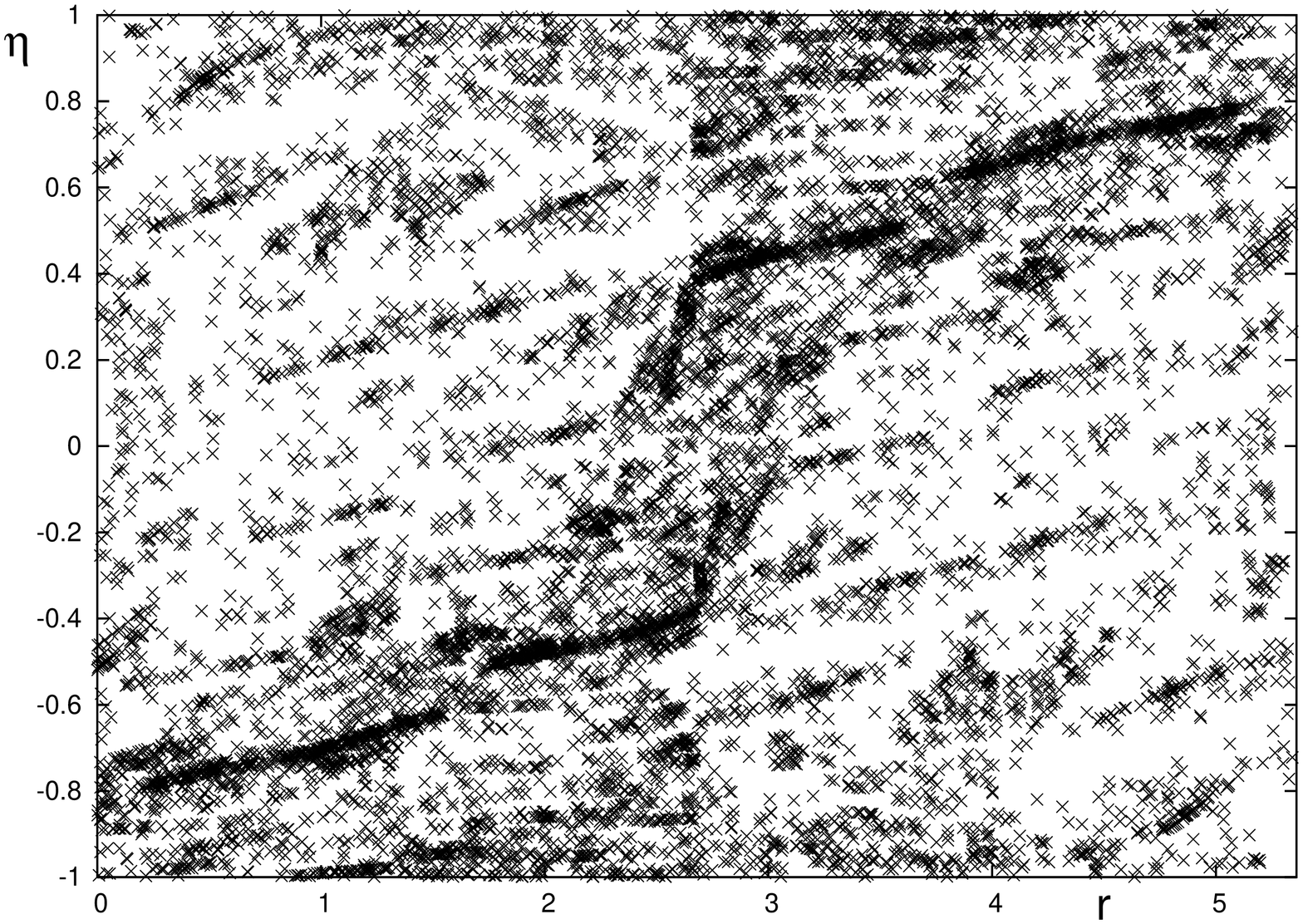,angle=-90,width=8cm}
\epsfig{file=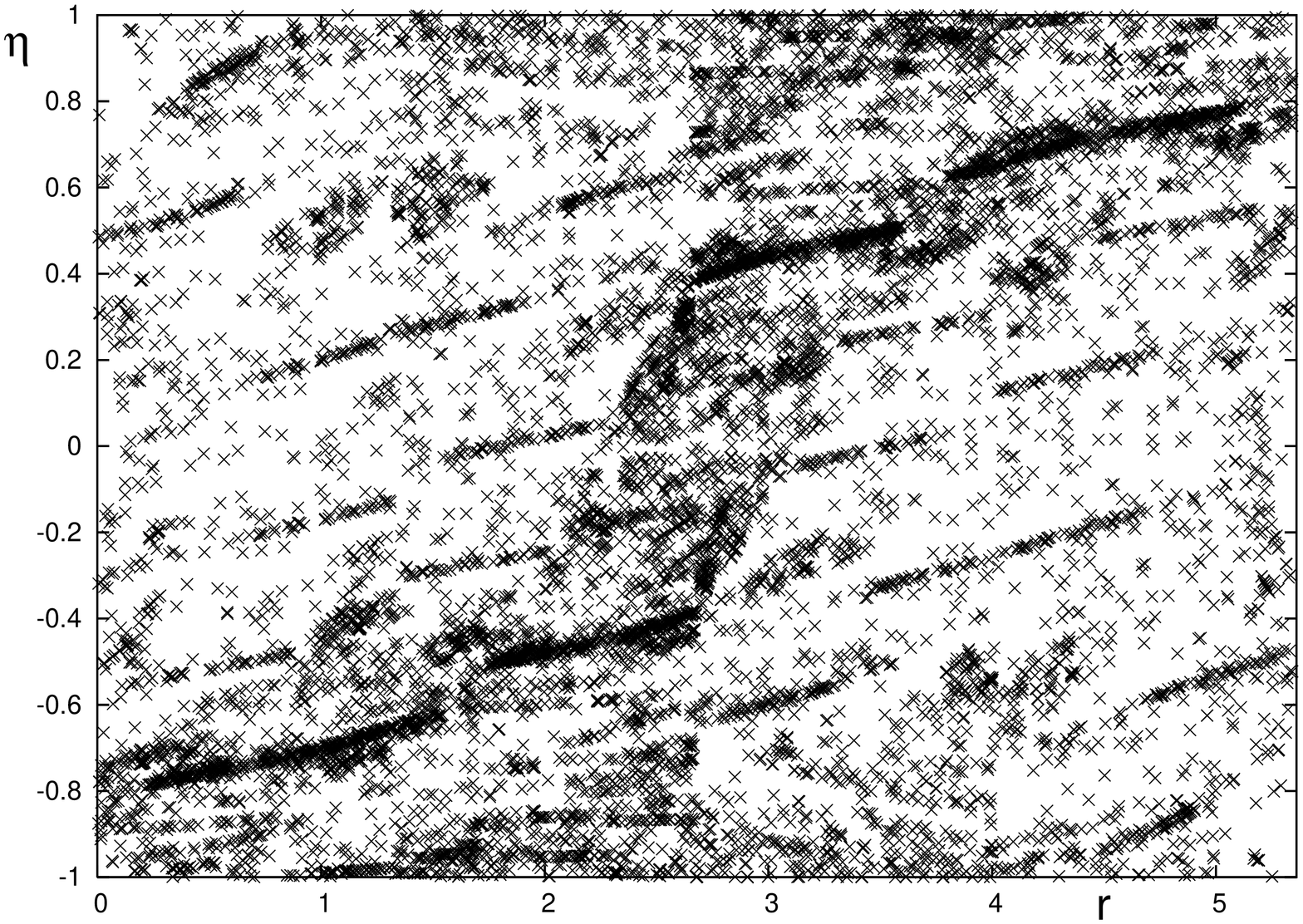,angle=-90,width=8cm}\caption{The last
$10^{4}$ iterates of the bounce map for $\epsilon=0.374$ out of a
trajectory of $5\times 10^{7}$ collisions (left panel) and for a
trajectory of $1.5 \times 10^{8}$ collisions, starting from the
last phase space point of the previous trajectory  (right panel).
The distribution appears to be the same,
suggesting that the system is in an apparently stationary state.
Here, $\eta = \cos \varphi$, with $\varphi$ the angle between the outgoing
velocity and the side of the rhombus, and $r$ is the perimetral distance
of the collision point from the right corner of the rhombus ($\eta$ and $r$
are called Birkhoff coordinates).}
\label{fig4}
\end{center}
\end{figure}

\begin{figure}
\begin{center}
\epsfig{file=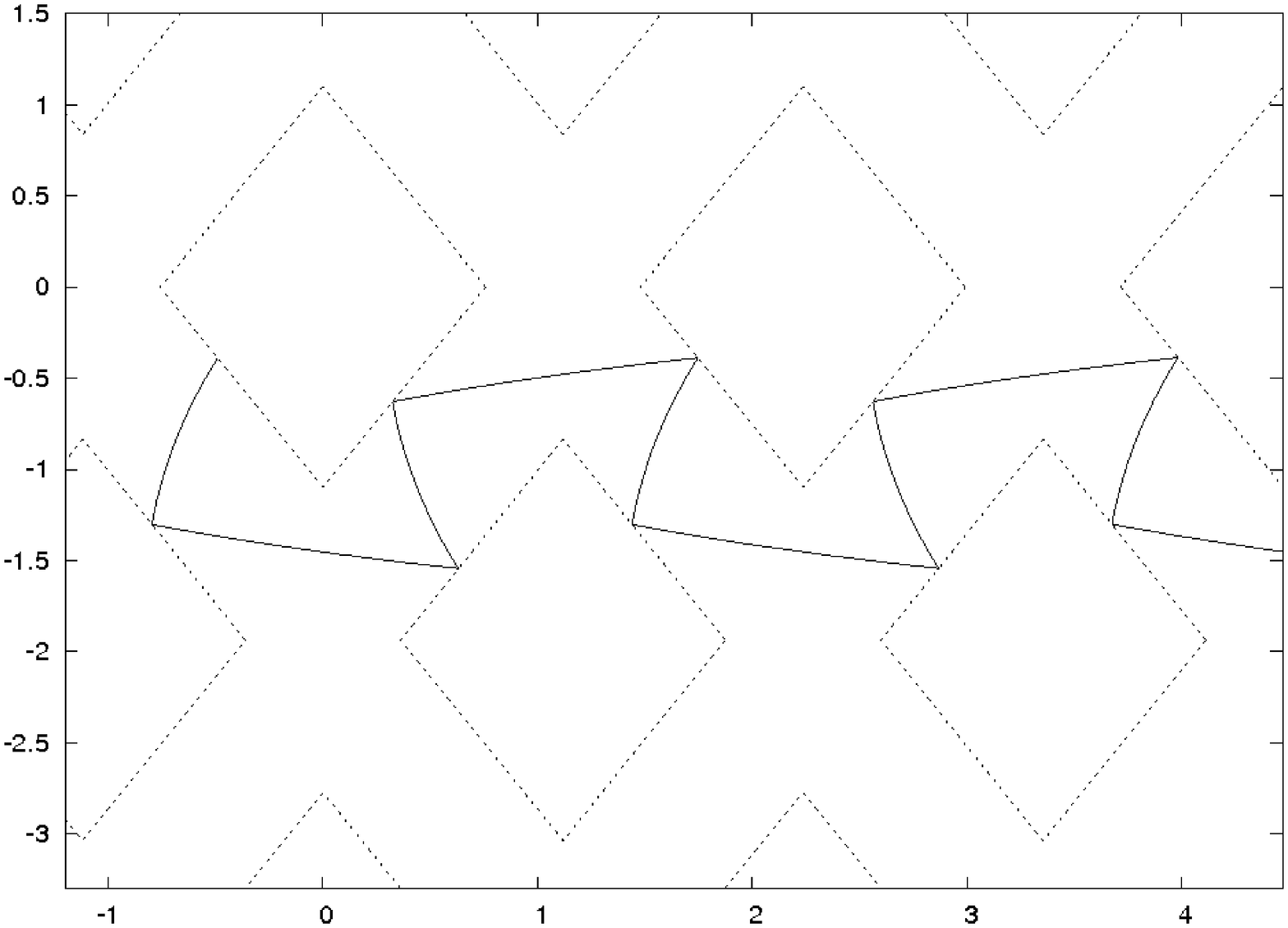,angle=-90,width=8cm}
\epsfig{file=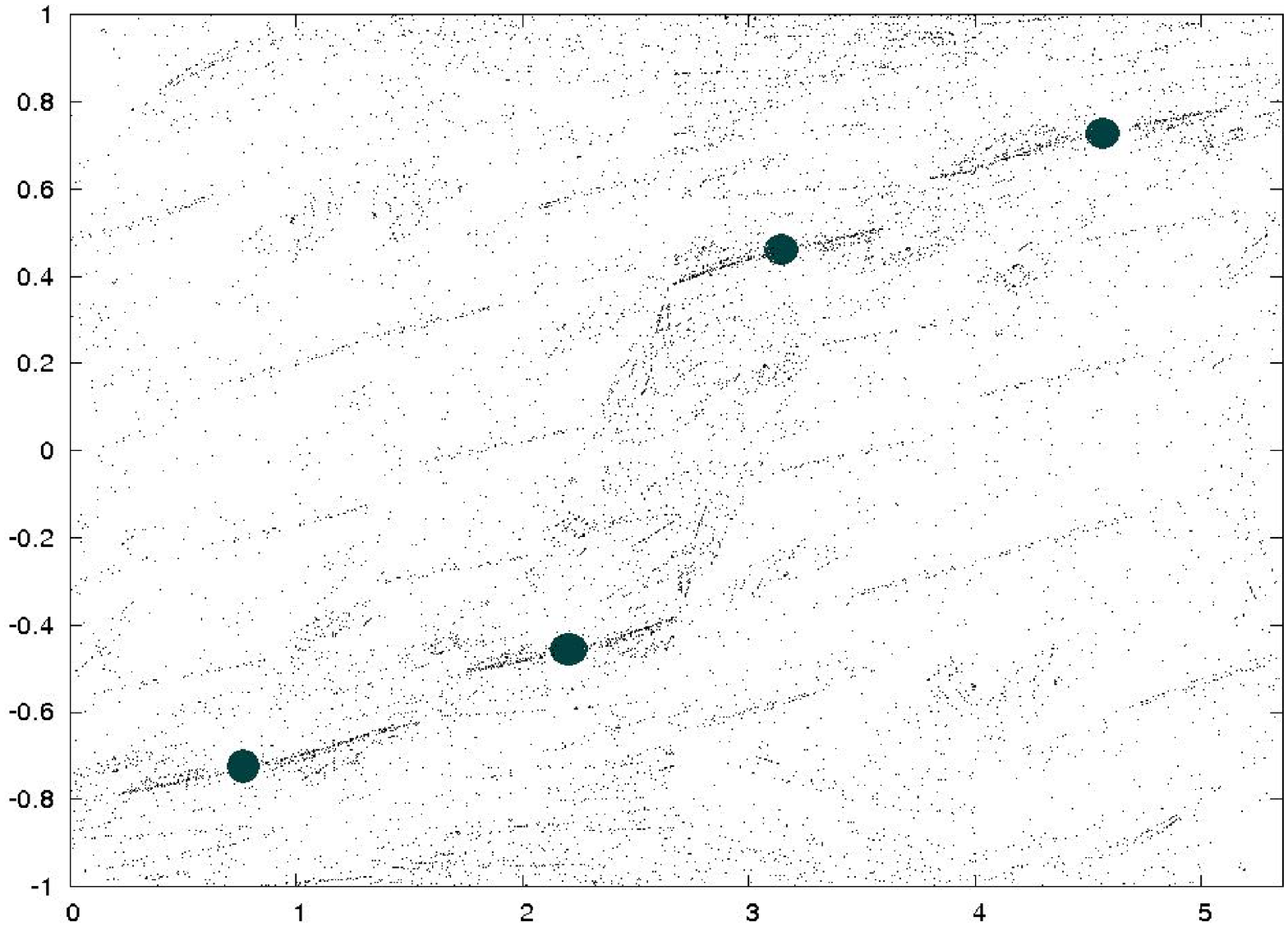,angle=-90,width=8cm}
\caption{The open period-four orbit
$(\mathcal{R}_{0,0},r_{3})(\mathcal{R}_{-1,-1}, r_{1})(\mathcal{R}_{1,-1}, r_{4})
(\mathcal{R}_{0,0}, r_{2})(\mathcal{R}_{2,0}, r_{3}) $
in the plane
for $\epsilon=0.374$,
$s_{y}=0.7573$, $s_{x}=1.1$, $L=1.291$, $\theta_{0}=-2.0619$,
$x_{0}=-0.4885$ (left panel).
The four points of this periodic orbit, are inflated to show their
embedding in the attractor (right panel). The coordinates are as in
Figure 4.
}
\label{fig7}
\end{center}
\end{figure}

\begin{figure}
\begin{center}
\epsfig{file=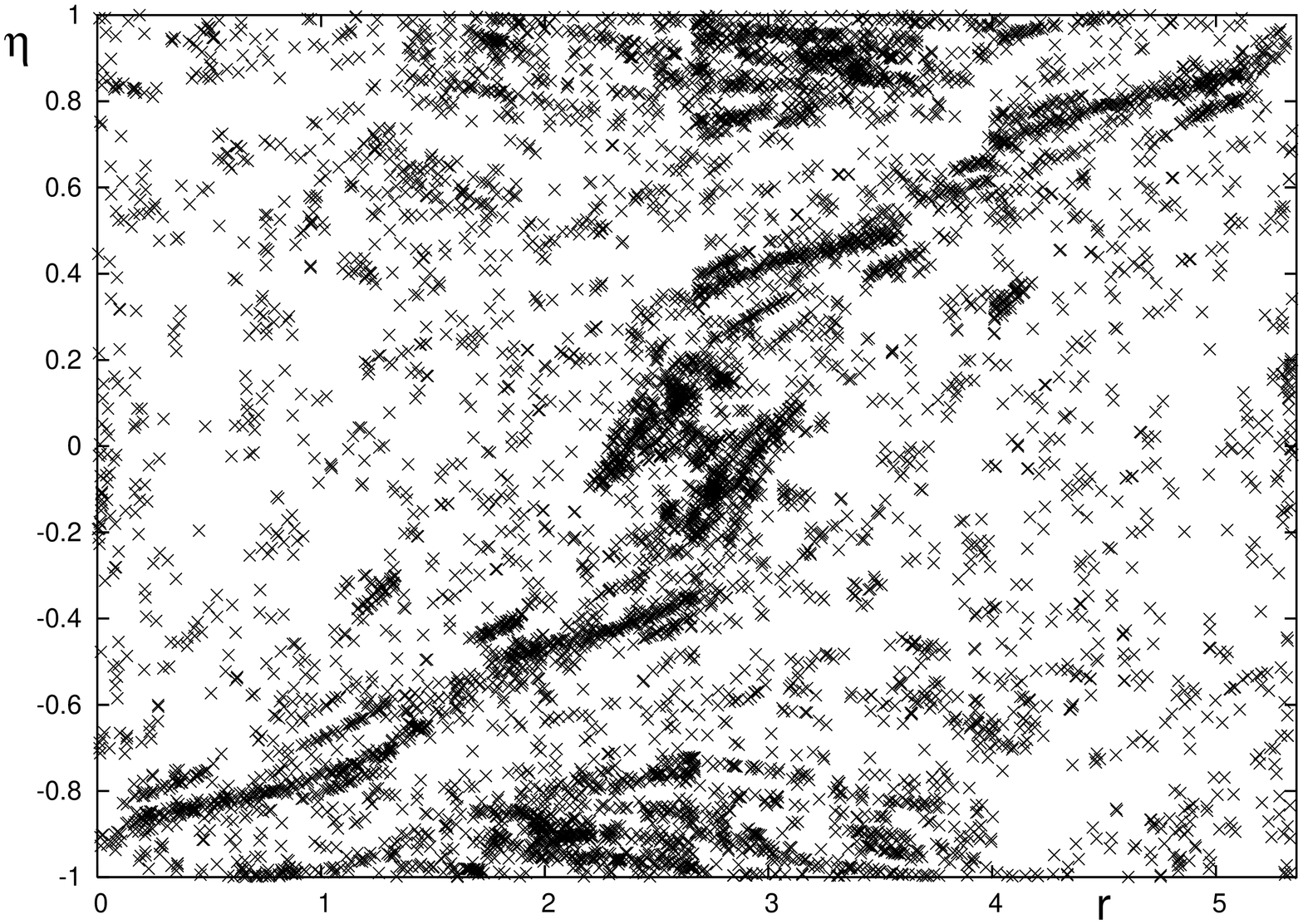,angle=-90,width=8cm}
\epsfig{file=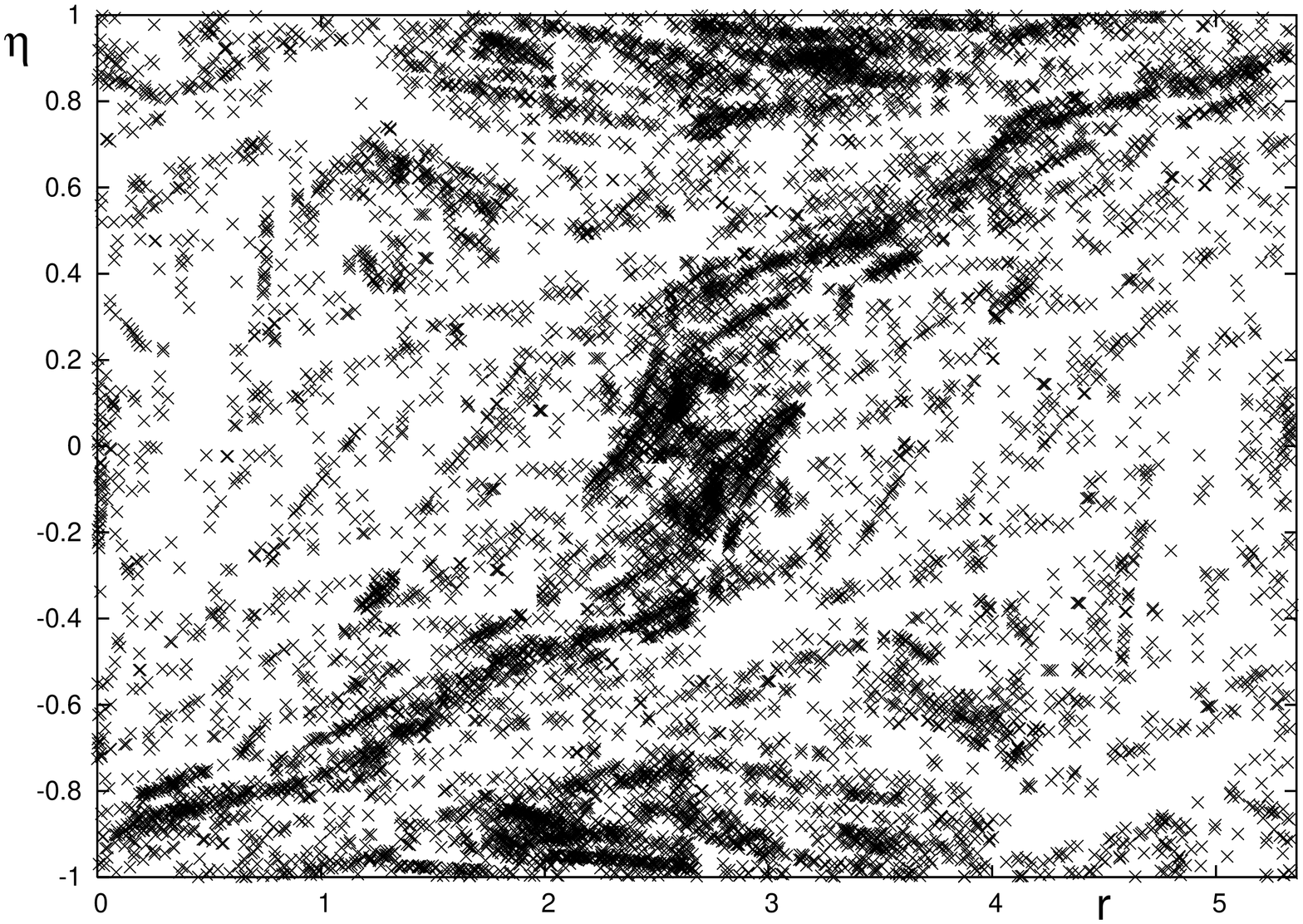,angle=-90,width=8cm}
\caption{The last
7000 points of the bounce map for $\epsilon=0.5$ out of a trajectory of length $5 \times 10^{7}$ collisions
 (left panel) and the last
$10^{4}$ points of a trajectory of length $2 \times 10^{8}$ with same
$\epsilon$ (right panel). The coordinates are defined as in
Figure 4.} \label{fig8}
\end{center}
\end{figure}

Therefore, we extended the simulations of the cases with
one apparently positive Lyapunov exponent, up to $1.5 \times
10^{8}$ collisions. The behaviour of the system still appears to be non
trivial, in cases such as that of
$\epsilon=0.374$. Furthermore, plotting the last $10^{4}$ iterates of
a trajectory of length $5 \times 10^{7}$ collisions, we
cover a large fraction of the phase space, which appears quite
close to that covered by the last $10^4$ iterates of a trajectory of $1.5 \times
10^{8}$ collisions (\fig{fig4}).\\
The conclusion that a chaotic stationary state has been reached
seems reasonable in this case, as the
computed positive Lyapunov exponent also indicates, having apparently converged to $0.144$
with three digits of accuracy,
after only $10^{6}$ collisions.\\
To strengthen this results, we have looked for an unstable periodic orbit embedded in the attractor, and
we have found one periodic orbit of period four, which apparently lies in the attractor and has
a positive Lyapunov exponent (\fig{fig7}, right panel). The other possibility is that the orbit
is isolated, and is separated from the attractor by such a small neighborhood that
is numerically impossible to resolve.\\
Another interesting example is given by $\epsilon=0.5$: the last
$7000$ points of a trajectory of length $5 \times 10^{7}$
collisions are compared with the last $10^{4}$ points
of the trajectory of length $2 \times 10^{8}$ (\fig{fig8}). Also
in this case the stationary state seems to have been reached, and
a periodic orbit of period four with one positive Lyapunov
exponent seems to be embedded in the attractor, similarly to the case of $\epsilon=0.374$.
The values of $\epsilon$, of the initial conditions and the Lyapunov exponents are reported in table I and II.
%
%
%
%
%
%

%
%
%
\begin{table}[ht]
\label{par-flat}
\begin{small}
\centering
\begin{tabular}{|c|c|c|}
\hline
Field & Collisions & Lyapunov exponent\\
\hline
0.374 & $10^{6}$ & 0.144232  \\
\hline
0.374 & $1.5 \times 10^{7}$ & 0.144317 \\
\hline
0.374 & $6.5 \times 10^{7}$ & 0.144291 \\
\hline
0.374 & $2.15 \times 10^{8}$ & 0.144320 \\
\hline
0.5 & $5 \times 10^{7}$ & 0.166648  \\
\hline
0.5 & $2 \times 10^{8}$ & 0.166622\\
\hline
\end{tabular}
\caption{Numerically computed Lyapunov exponents for different trajectories and fields.}
\end{small}
\end{table}
\begin{table}
  \centering
  \begin{tabular}{|c|c|c|c|c||c|}
\hline
Field & $x_{0}$ & $y_{0}$ & $\theta_{0}$ & $\lambda_{1}$ &  $\lambda_{2}$ \\
\hline
0.374 & $-0.48971578282741385$ & $-0.3886737605834475$ & $-2.06199461833$ &$0.108316$&$-0.281452$\\
\hline
0.5 & $0.30576674801010$ & $0.6558650167554337$ & $-0.31873995693500$ & $0.155569$& $-0.384674$\\
\hline
  \end{tabular}
  \caption{Initial conditions and Lyapunov exponents of periodic orbits of period four. All data
have been computed analitically.}
  \label{par-flat1}
\end{table}
\subsection{Chaos for small electric fields}

\begin{figure}
\begin{center}
\epsfig{file=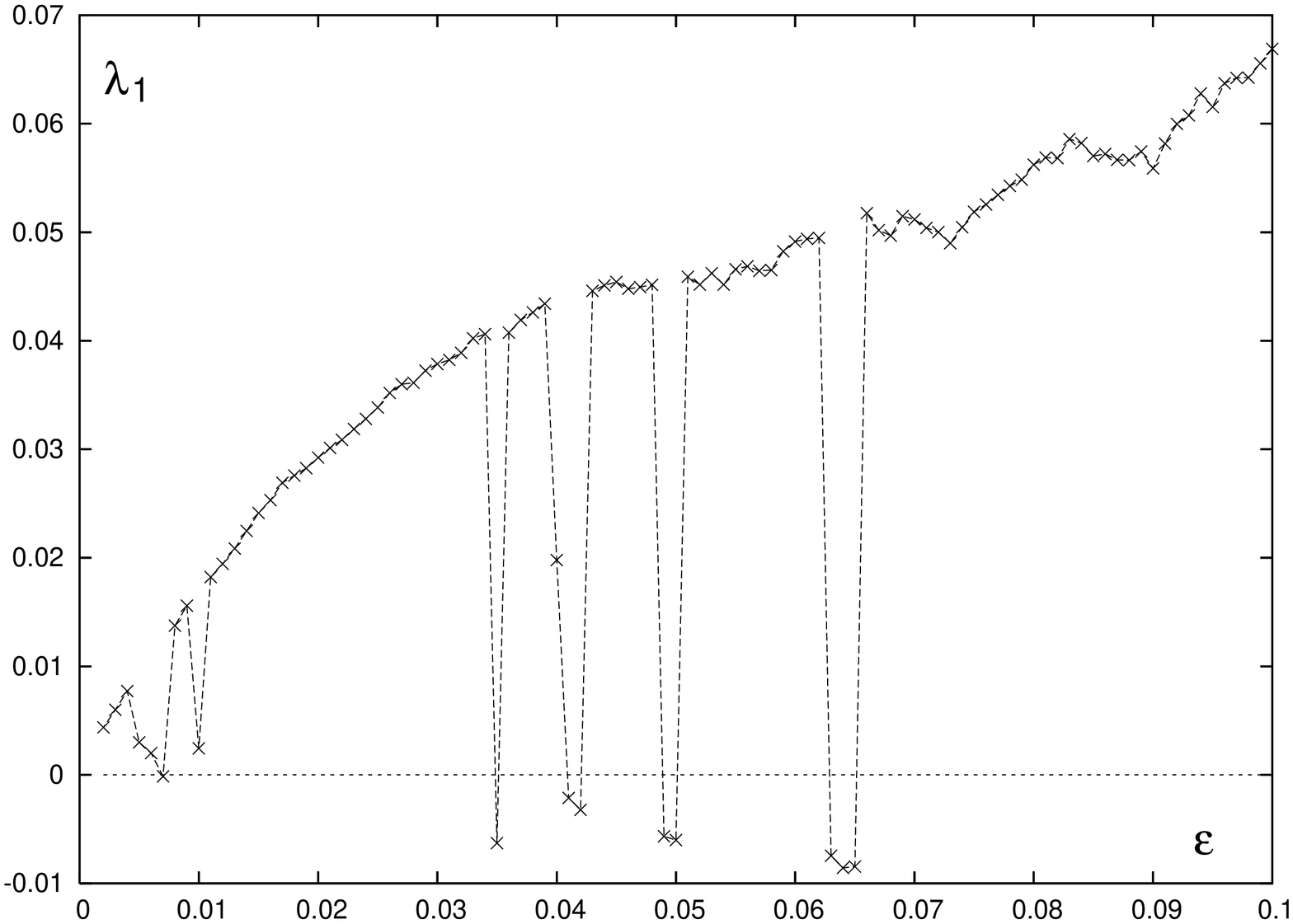,angle=-90,width=8cm}
\epsfig{file=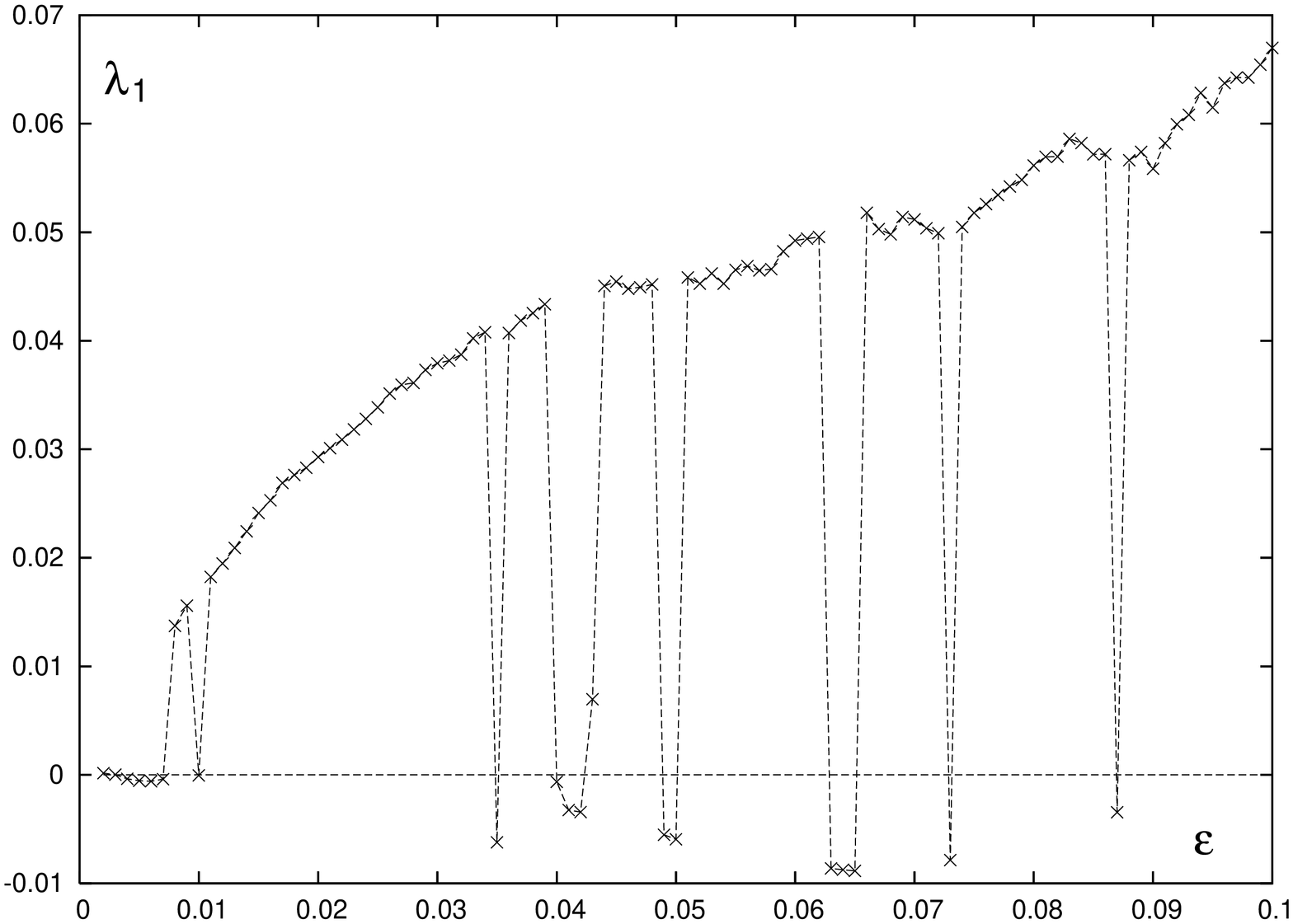,angle=-90,width=8cm}
\caption{
The largest Lyapunov exponent, $\lambda_1$, for different lengths of trajectory:
$10^{7}$ (left panel) and $3 \times 10^{8} $ (right
panel) for $\epsilon \in [0.002,0.1]$ and random initial condition.} \label{fi1}
\end{center}
\end{figure}

Numerical simulations of the model were performed starting with
random initial conditions and considering the electric field in the
interval $[0.002, 0.1]$, for trajectories of length $n=10^{7}$.
Looking at \fig{fi1} (left panel) we find more cases with one
apparently positive Lyapunov exponent than with two negative
exponents.
Plotting the last $10^{4}$ points, in trajectories of length
$1.5\times 10^{8}$, for the electric fields that produced
apparently positive Lyapunov exponents,
the situation is pratically the same as for large fields:
some cases with one exponent that appeared to be
positive after $10^{7}$ collisions eventually produce (after $1.5 \times 10^{8}$ collisions) two negative
exponents and a periodic or
quasi-periodic steady state \textbf{\cite{LRB}
\footnote{as in \textbf{\cite{LRB}}, it is always
difficult to decide whether these orbits are periodic or
quasi-periodic}}.
We have further increased the number of collisions  up
to $3 \times 10^{8}$, but the exponent $\lambda_1$
remained positive in most of the cases (\fig{fi1}, right panel) and an apparently chaotic attractor was reached.
%
%
%
%
%
%
%
%
%
%
\subsection{A small basin of attraction}

\begin{figure}
\begin{center}
\epsfig{file=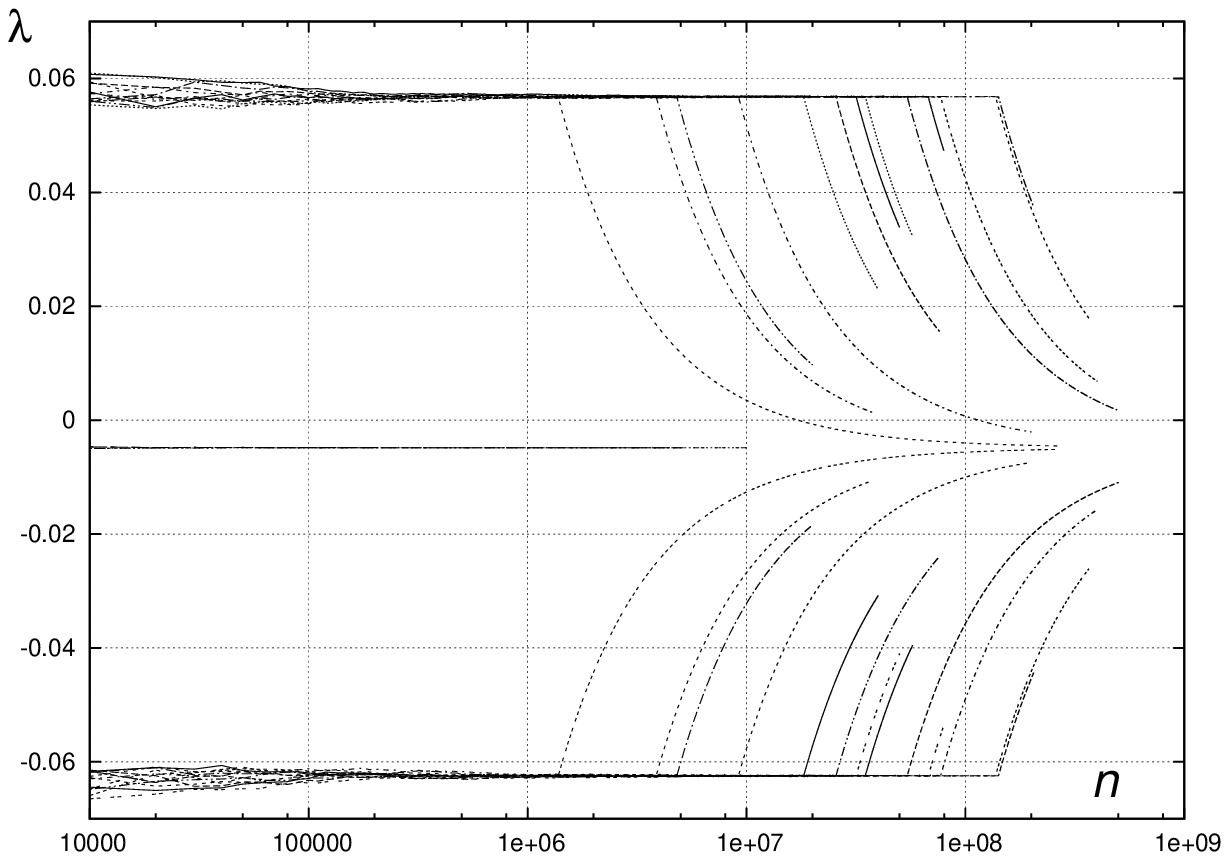,angle=0,width=8cm}
\epsfig{file=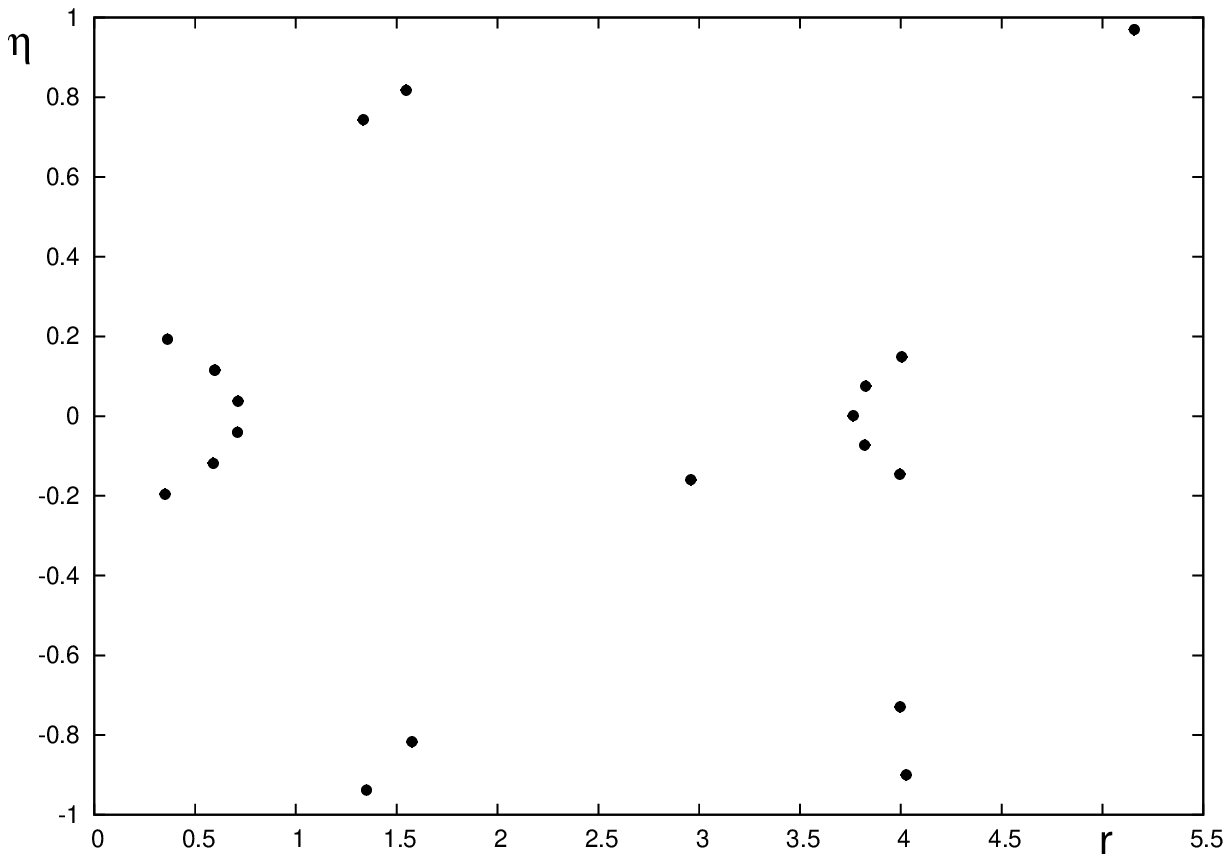,angle=0,width=8cm}
\caption{Behaviour of the finite time Lyapunov exponents $\lambda_{1,2}$ with the
number of collisions $n$, for $\epsilon=0.087$ (left panel).
The logarithmic scale in $n$ clearly separates
the initial, intermediate and asymptotic regimes. The
exponents rapidly converge to one positive and one negative value, which persist
for very long, up to a number $N_c$ of collisions, dependent on the initial
condition.
After $N_c$, both exponents converge as $1/n$ to the value $-0.004838$.
For random initial conditions, $N_c$ is of order $O(10^8)$, for initial conditions
close to the asymptotic periodic orbit (right panel), $N_c$ varies irregularly between
$O(10)$ and $O(10^7)$. The right panel reports the last $10^4$ points of
$3\cdot10^8$ collisions, which testifies that the motion has
settled on a periodic orbit of 19 points only. The coordinates are defined
as in Figure 4.} \label{fii1}
\end{center}
\end{figure}

The behaviour illustrated above is rather peculiar and calls for some explanation.
How can it be that a steady
state is so hard to reach in so many cases? Usually, convergence to an attracting
periodic orbit occurs rather quickly, while doubts remain in some of the cases
we considered even after $10^8$ iterations of the bounce map. Therefore, we
have investigated in greater detail the specific example with
$\epsilon=0.087$, $L=1.291$, $s_x = 0.7573$ and $s_y = 1.1$.
One finds that the largest time dependent Lyapunov exponent, $\lambda_{1}$,
rapidly settles on a positive value, as if the trajectory had
reached a chaotic attractor. However, for randomly chosen initial conditions,
a striking and precise monotonic $1/n$ behavior sets in for $\lambda_1$, after
a critical, typically large, time $N_{c}$ (cf.\ left panel of \fig{fii1}), as if
the trajectory had eventually collapsed on an attracting  periodic orbit.
Indeed, the asymptotic value of $\lambda_1$ is $-0.004838$ with an estimated error
not larger than $10^{-6}$ \textbf{\footnote{Incidentally, as common in nonequilibrium billiards
\textbf{\cite{Ro3}}, the second Lyapunov exponent takes the same value
as the first, because the eigenvalues of the stability matrix are
complex conjugate.}}. The accuracy of this result is due to the fact that a
sufficiently long simulation approaches an attracting periodic orbit to
practically full numerical precision; the Lyapunov exponents can then be
computed with analogous accuracy. This is particularly true in the present
example, which turns out to have an attracting orbit made of only 19 points,
whose initial condition, up to 12 digits accuracy, is given by
$x_{\rm po}=0.418447478686, y_{\rm po}=0.492193019206,\theta_{\rm po}=0.718794450586$,
(cf.\ right panel of \fig{fii1}).
For such a small number of points, numerical errors cannot appreciably affect
the result. In particular, no doubts remain, in this case, about the negative
sign of both exponents, hence about the attracting nature of the orbit.

The question now arises as to the shape and size of the basin of attraction
$\mathcal{B}$ of the asymptotic periodic orbit, because a particle with random
initial conditions wonders around almost all phase space before falling inside
this set.

Our analysis of the evolution of trajectories, with initial condition close to
the attracting periodic orbit, shows that $\mathcal{B}$ is quite limited in
size, and particularly hard to reach because it contains a very small region
around the right vertex of the rhombus. Furthermore, by varying the initial
conditions around $(x_{\rm po}, y_{\rm po},\theta_{\rm po})$, and computing
the Lyapunov exponents, the irregular shape of $\mathcal{B}$ is evidenced by
the times $N_c$, which vary most irregularly from $O(10)$ to $O(10^8)$,
$O(10^8)$ being typical for random initial conditions.
Nevertheless, the {\em radius} of $\mathcal{B}$, i.e.\ the supremum distance
between any two points of $\mathcal{B}$, is not smaller than
$10^{-4}$, which is quite small but well above the distances which can be
accurtely measured with double precision numerical simulations.

We conclude that $\mathcal{B}$ lies at the border of a chaotic repeller
\textbf{\cite{TCTel}}, and that
it is quite small and of irregular shape. Hence, only after a sufficiently
long peregrination in phase space, when the transiently chaotic trajectory gets
sufficiently close to $\mathcal{B}$, may numerical errors let the trajectory jump
inside $\mathcal{B}$. At this stage, the sudden (exponential) convergence
of the trajectory and the consequent $1/n$ behaviour of the
Lyapunov exponents begin.

The same qualitative behaviour has been observed for $\epsilon=$0.002,
0.003,
0.004,
0.005,
0.006,
0.007,
0.010,
0.022,
0.040,
0.041,
0.042,
0.043,
0.050,
0.064,
0.065,
0.073,
0.087.
%
%
%
%
%

\section{Behaviour of the attractor}

It is interesting to understand the behaviuor of the steady state
with the field, i.e. to build a kind of bifurcation diagram, e.g.\
to compare with the one for the Lorentz gas, whose
obstacles are defocussing \textbf{\cite{Ro3}}. Our analysis
reveals substantial differences from the case of \textbf{\cite{Ro3}},
as well as from standard low dimensional dynamics.

\subsection{Multi-furcation as function of the electric field}

\begin{figure}[ht]
\begin{center}
\epsfig{file=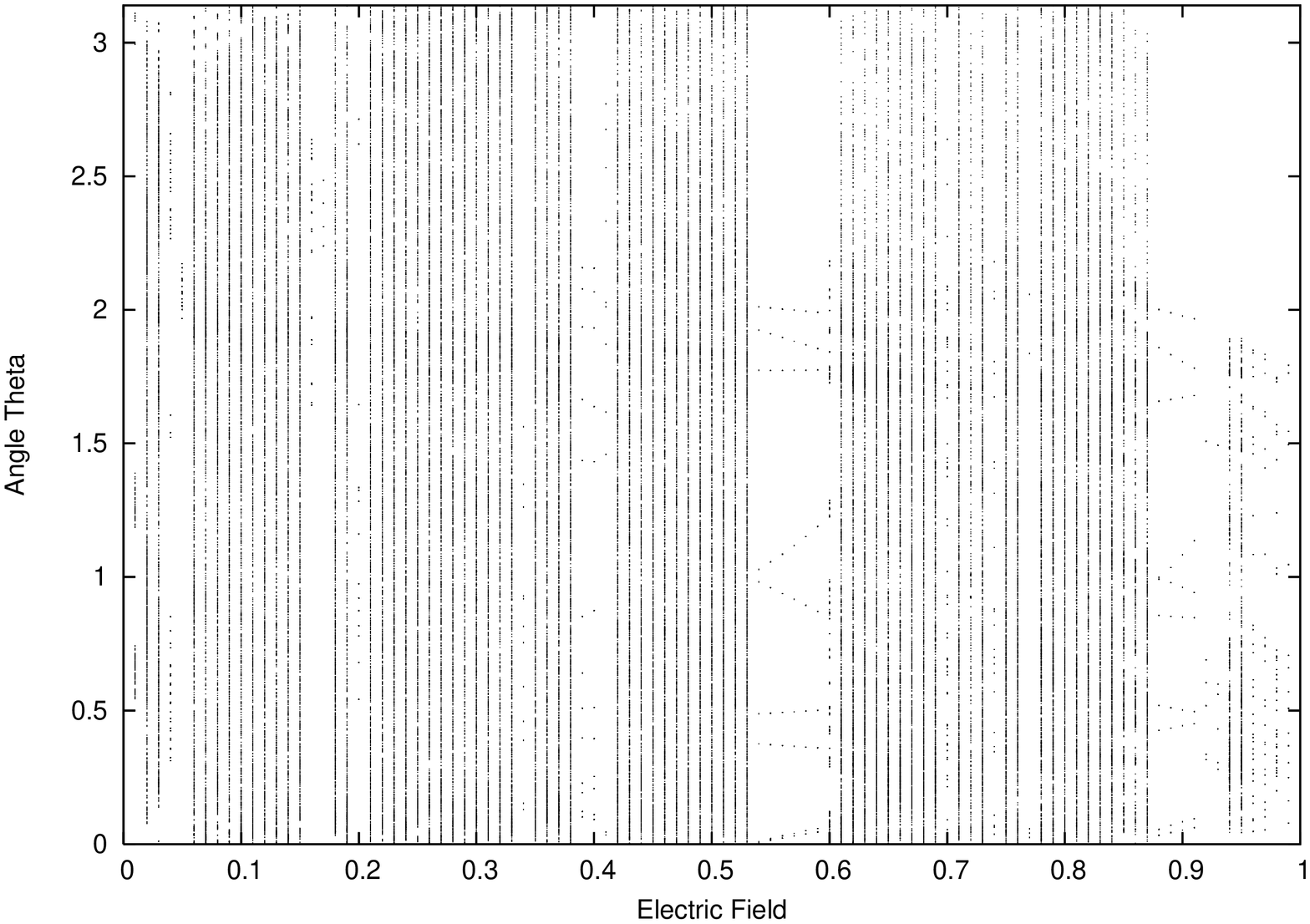,angle=-90,width=8cm}
\epsfig{file=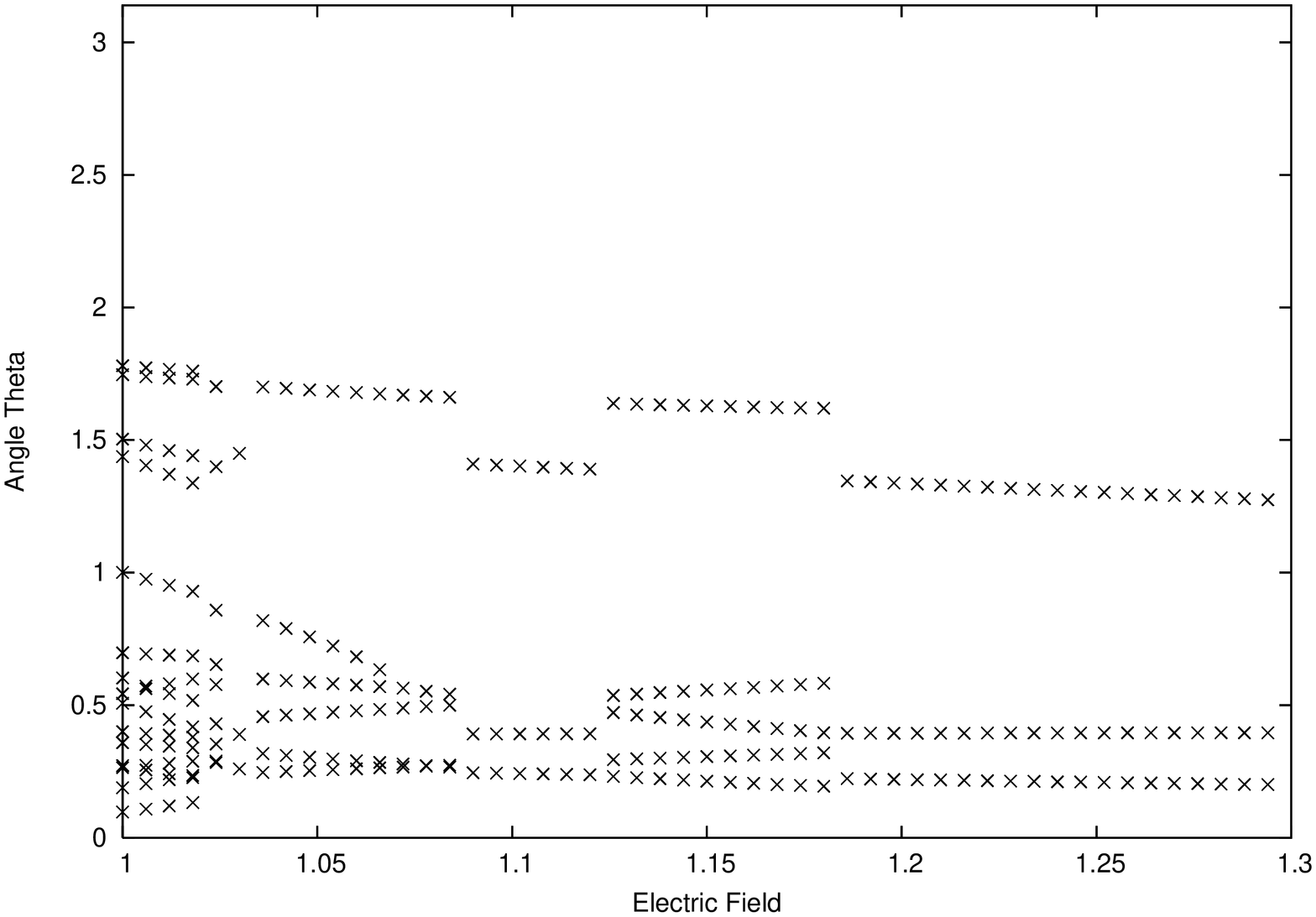,angle=-90,width=8cm}
\caption{The upper
half of the projection of the multi-furcation diagram onto the $\theta$ axis,
for $\epsilon $ in the range $[0.01;1]$. The last
$10^{3}$ points of trajectories with length $10^{7}$ collisions have been plotted (left panel). The
multi-furcation diagram in the $(\epsilon;\theta)$ plane, for
$\epsilon$ in the range $[1,1.3]$, from the last $10^{3}$ collisions out of a
trajectory of $10^{7}$ collisions (right panel). }
\label{bif001-1}
\end{center}
\end{figure}

In order to visualize the behaviour of the attractor as a
function of the electric field, we consider a projection of the
billiard map phase space: the projection onto the $\theta$ axis, which shows a sort of ``multi-furcation'' scenario. When the field is varied, a series of dramatic
changes in the dynamics occurs.\\
For the geometry determined by $L=1.291$, $s_{x}=0.7573$ and
$s_{y}=1.1$, numerical simulations were performed for a random initial
condition, ignoring the initial transient behaviour, for the
electric fields in the range $[0.01,1]$ with a step
$\Delta\epsilon=0.01$. At the end of a trajectory of $10^{7}$
collisions, the last $10^{3}$ points were plotted (\fig{bif001-1},
presents only the upper half projection onto the $\theta$
axis; the other half
 of this diagram is trivially obtained from this, by reflection along the line $\theta=0$, as a
 result of the symmetry imposed on the system by the external field). In this way, many electric fields are found whose
dynamics sample most of the $\theta$ space, while only a few
fields show a periodic (or quasi periodic) steady state. The
apparently chaotic regions and the regular regions
are finely interspersed with each other, in a way which we have not found elsewhere in the literature.\\
We have performed simulations also in the range $[1,1.3]$ and have only plotted the
 last $10^{3}$ points out of $10^{7}$ collisions, as in the previous case.
 In this range, all attractors have been found to be  periodic orbits, which are reached well
 before $10^{7}$ collisions. In other words, the stationary state
 is rapidly reached with large electric fields as expected for the correspondingly high dissipations.\\
How conclusive are these results? Again, when a periodic steady
state is reached, the situation is clear, while doubts remain when the steady
state looks chaotic. Therefore we have considered the range $[0.77,0.86]$:
a rather wide range, apparently chaotic, but also characterized by high
dissipation, which favours ordered dynamics.
Running trajectories of $10^{8}$
collisions and plotting the last $10^{3}$ points, the apparently weakly chaotic
states survived, despite the high dissipation. This study combined with the
analysis of the largest
Lyapunov exponent makes chaos quite plausible for these fields, although we cannot exclude that the trajectories collapse on periodic orbits after much longer times. On the other hand, there is no purpose in pushing further this analysis, since it is bound to remain uncertain. Here, it suffices to have uncovered a rather peculiar behaviour, unexpected for simple dynamical systems.\\
Indeed, in our model, the passage from low-period attractors to apparently
chaotic steady states, is rather abrupt and seems to be discontinuous.
Moreover, the periodic attractors always seem to coexist with transiently
chaotic states. Differently, standard bifurcation scenarios are characterized
by a gradual growth of the period of the attracting orbits, the orbits are {\em globally}
attracting and lead to fast convergence of the trajectories towards them.
To further investigate this behaviour, we have honed the range $[0.77,0.78]$
with step $0.001$ simulating trajectories of $10^{8}$ collisions, finding a (quasi) periodic orbit at $\epsilon=0.771$ and another one at $\epsilon=0.777$, with apparently chaotic trajectories in the middle.
Thus, we have further honed the range $[0.770;0.771]$ with a step of $0.0001$, to find
that a jump from (quasi) periodic orbits to apparent chaos happens for a field
variation of just $10^{-5}$.\\
We have also performed simulations in the range $[0.771,0.772]$, namely
from the (quasi) periodic orbit at $\epsilon=0.771$ to an apparently
chaotic case at $\epsilon=0.772$. Plotting the last $10^{3}$ points of trajectories of
$10^{8}$ collisions, apparent chaos and (quasi) periodic orbits appear again to be finely
intertwined. The range $[0.18,0.38]$ was similarly studied, and some of the cases that appeared to be chaotic after $10^{7}$ collisions, turned into
(quasi) periodic after $10^{8}$, but not all. Therefore, the duration of the transients
results exceedingly long in all cases, which is a manifestation of either very small
basins of attraction, or of the smallness of the stable islands,
whose size would then vary wildly with the field. Although this is not conclusive evidence, our analysis supports the idea that
there could be a discontinuous transition
between chaos and regular motion, in the non-equilibrium Ehrenfest gas.

\subsection{Dependence on the geometry}

In this section we outline the behaviour of the attractors as
functions of one of the parameters  which determine the shape of
the billiard: $s_{x}$. We take two electric fields (one large
and one small) for which we had found apparently chaotic behaviours.\\
Let
$\epsilon=0.374$ be the electric field,  $L=1.291$ and
$s_{y}=1.1$; we study the attractor reliance on $s_{x} \in [0,1]$ with
step $0.01$. We considered
trajectories of length $10^{7}$ collisions and looked at the last
$10^{3}$ points. We found a complicated scenario, analogous to the one for
the dependence on $\epsilon$, as $s_{x}$ was varied. In the
chosen range, there are more cases with apparently chaotic attractors than
with (quasi) periodic steady states. Increasing the number of collisions
in each trajectory up to $5 \times 10^{7}$, we find that apparent chaos
persists, in other words it seems that
the effect of the electric field prevails on the geometry effects.
To analyze more carefully this fact, we have honed the interval
$[0.09,0.1]$ for $s_{x}$, with step $0.0005$ and for trajectories of
$10^{8}$ collisions. A sudden transition between apparent chaos
and periodic orbits happens at $s_{x}=0.099$.\\
%
%
%
%
For $\epsilon=0.014$,  $s_{x}$ was taken in $[0,1]$  with step
$0.01$, for trajectories of length $10^{7}$ collisions. We found that the situation is
slightly different from the small field case. The apparently chaotic steady states
occupy a smaller region of the phase space than the steady
states of the  $\epsilon=0.374$ case. For instance, the
asymptotic state of $\epsilon=0.014$ seems to fill almost
completely the space $\theta$, differently from the case of
$\epsilon=0.374$. Increasing the length of the trajectories up to $5 \times 10^{7}$ collisions, various apparently chaotic cases reduce to
(quasi) periodic orbits, while
other survive. Finally, we increased the number of collisions up to
$10^{8}$ for $\epsilon=0.014$ again finding that some apparently chaotic cases turned into quasi periodic cases, while other survived. \\
We conclude that the dependence of the attractors on the parameter
$s_x$ is qualitatively similar to the dependence on $\epsilon$, although
at times the dependence on $\epsilon$ prevails.
%
%
%
%
\section{Conclusion}

In this paper we have examined the non equilibrium version of the
Ehrenfest gas, which is a billiard model with an electric field and a Gaussian
thermostat, whose point particle moves in
the plane and undergoes elastic collisions with rhomboidal
obstacles. The motivation was to understand the dynamics of this
nonequilibrium model, whose obstacles have flat surfaces, hence do not
defocus the trajectories, while its thermostat, which makes dissipative the
dynamics, does focus them \textbf{\footnote{The dissipation produced by the Gaussian thermostat
sharply differentiates our dynamics from Hamiltonian dynamics, although both
preserve the total energy. Indeed, in our case, the total energy equals the kinetic
energy, which is a constant of motion.}}. We have shown that periodic
orbits with one positive Lyapunov exponent embedded in what appear to be
chaotic attractors, do exist. Our
numerical results have identified electric fields whose
dynamics is strongly suggested to be chaotic, although conclusive results are out of reach at present, because of a very
peculiar phenomena. The stationary state, even if attracting and trivial,
requires very long times to be reached, because it coexists with a
transiently chaotic state which covers most of the phase space. The
dependence on the model parameters of the steady
state behaviour  also presents peculiar features which, to the best of our knowledge, have not been observed before: a very irregular, possibly discontinuous dependence of the attracting orbit, and/or of the size and shape of the stable islands,
on $\epsilon$ and $s_{x}$. The peculiarity remains even if it is
eventually proved that all steady states are periodic, because of their
coexistence with transiently chaotic states, and because of the consequent
irregular behaviour of the convergence rates.
As the sum of the Lyapunov exponents
is proportional to the travelled distance, cf.\ Eq.(\ref{formula}), very irregular
transport properties are obtained as well, perhaps as irregular as those
conjectured for other low-dimensional dynamical systems \textbf{\cite{kla}}.
\section{Acknowledgements}
\label{ack}

We would like to thank M. P.~Wojtkowski for his useful comments on a preliminary version of this paper and for enlightening discussions. We are indebted to O.G~Jepps and C.M.~Monasterio for  continuing
encouragement and suggestions. LR is grateful to ESI and to the organizers of the ESI Semester
on "Hyperbolic Dynamical Systems", May 25 - July 6, 2008, where part
of this work has been developed. CB has been partly funded by the Lagrange Foundation, Torino, Italy.

\end{document}